\documentclass[a4paper,11pt]{article}
\pdfoutput=1 % if your are submitting a pdflatex (i.e. if you have
\usepackage{jheppub}

\usepackage{graphicx, epstopdf}

\usepackage[utf8]{inputenc}

\usepackage{amsmath, amsthm, amsfonts, amssymb}
\usepackage{mathtools, mathabx, bm, bbm, scalerel, xfrac}
\usepackage{empheq, physics, slashed, cancel}
\usepackage[usenames,dvipsnames]{xcolor}
\usepackage{calc}
\usepackage{comment}

\usepackage{hyperref}
\usepackage{color}
\usepackage{subcaption}

\newcommand{\D}{\text{d}}

\newcommand{\tb}{\bar{t}}
\newcommand{\kb}{\bar{k}}
\newcommand{\qb}{\bar{q}}

\newcommand{\mts}{m_t^{\text{small}}}
\newcommand{\lo}{(0)}
\newcommand{\nlo}{(1)}
\newcommand{\msb}{\overline{\text{MS}}}
\newcommand{\Qt}{\tilde{Q}}
\newcommand{\M}{|\mathcal{M}|^2}

\newcommand{\tightoverset}[2]{%
  \mathrel{%
    \raisebox{1.1ex}{%
      \scalebox{0.6}{%
        $\scriptstyle #1$%
      }%
    }%
    \mkern-11.5mu%
    #2%
  }%
}

\newcommand\qbm{\tightoverset{\text{\tiny(\hspace{6px})}}{\qb}}

\title{Top-associated Higgs-boson production using perturbative fragmentation functions at next-to-leading-order}

\author[a]{Colomba Brancaccio,}
\author[b]{Micha\l{} Czakon,}
\author[c]{Terry Generet,}
\author[b]{Benedikt Gurdon}
\affiliation[a]{Dipartimento di Fisica and Arnold-Regge Center, Università di Torino, and INFN, Sezione di Torino, Via P.\ Giuria 1, I-10125 Torino, Italy}
\affiliation[b]{Institut f\"ur Theoretische Teilchenphysik und Kosmologie, RWTH Aachen University, D-52056 Aachen, Germany}
\affiliation[c]{Cavendish Laboratory, University of Cambridge, Cambridge CB3 0HE, U.K.}

\emailAdd{colomba.brancaccio@unito.it}
\emailAdd{mczakon@physik.rwth-aachen.de}
\emailAdd{generet@hep.phy.cam.ac.uk}
\emailAdd{benedikt.gurdon@rwth-aachen.de}

\abstract{Under certain conditions, the production of a Higgs boson in association with a top-anti-top pair at hadron colliders can be described via a factorisation theorem using perturbative fragmentation functions. The latter describe the nearly collinear emission of a Higgs boson from a top-quark and reproduce the leading mass dependence of the exact next-to-leading-order (NLO) calculation. Although the NLO fragmentation functions have been calculated a few years ago, it has not been possible up to now to demonstrate the applicability of the approximation in a realistic setup. At NLO, we analyse two different ways of treating the top-quark mass, called the zero-mass-top-quark (ZMTQ) and the hybrid prescription. We show that the method yields reliable results at LHC center-of-mass (cms) energies in the hybrid prescription. In the ZMTQ prescription, the results at LHC cms energies are only reliable in the quark-anti-quark channel, but become viable for the full $pp \to t\bar{t}H$ process at a 100 TeV hadron collider. In addition, we discuss some subtleties and complications arising when extending the formalism to next-to-next-to-leading-order (NNLO) and beyond.}
\preprint{P3H-26-037, TTK-26-14, Cavendish-HEP-26/03}

\begin{document}
\maketitle
\flushbottom

\section{Introduction}

Higgs boson production in association with a top-quark pair plays an important role in searches for physics Beyond the Standard Model (BSM) and provides an opportunity to measure the top Yukawa coupling $y_t$ directly. By now, both ATLAS and CMS have measured this process using their respective run II data sets \cite{ATLAS:2024gth,CMS:2024fdo}. While ATLAS reports a total cross section in good agreement with the SM prediction, the CMS signal strength of $\mu=0.33\pm 0.26$ is about $2.5\,\sigma$ away from the expectation. Given the large errors for the total cross sections, no differential measurements exist beyond simplified template cross sections.

On the theory side, the first NLO calculations were completed more than 20 years ago \cite{Beenakker:2001rj,Reina:2001bc,Reina:2001sf,Beenakker:2002nc,Dawson:2003zu} and the NLO electro-weak (EW) corrections followed around 2015 \cite{Frixione:2014qaa,Zhang:2014gcy,Frixione:2015zaa,Frederix:2018nkq}. The first step towards next-to-next-to-leading order (NNLO) results were the flavour off-diagonal partonic channels calculated in Ref.~\cite{Catani:2021cbl}. Since then, full results have been published using a soft-Higgs approximation for the double virtual amplitudes \cite{Catani:2022mfv}. In addition, differential results were presented in Ref.~\cite{Devoto:2024nhl} which combines the soft-Higgs approximation with a high-energy expansion in the small top-mass limit. Progress regarding the exact two-loop corrections has been made in recent years, but no full calculation is available yet (see, e.g.\ Refs.~\cite{FebresCordero:2023pww,Agarwal:2024jyq,Wang:2024pmv}).

In addition to these higher-order effects, soft-gluon resummation results are available at next-to-next-to-leading logarithmic (NNLL) accuracy \cite{Kulesza:2015vda,Broggio:2015lya,Broggio:2016lfj,Kulesza:2017ukk,Kulesza:2018tqz,Broggio:2019ewu,Kulesza:2020nfh}.

All of these results have been combined in Ref.~\cite{Balsach:2025jcw} which matches the approximate NNLO calculation with the NNLL soft-gluon resummation and takes into account the full NLO effects (including all leading and subleading EW effects). Furthermore, NNLO results obtained using approximated double virtual amplitudes have recently been matched to a parton shower via the $\verb|MiNNLO|_{\text{PS}}$ method~\cite{Biello:2026nhj}.

In contrast to the soft-Higgs approximation where the Higgs momentum is assumed to be small, we analyse the behaviour when the Higgs is very hard, i.e.\ the high tails of the Higgs transverse momentum, $p_{T,H}$, distribution. At $p_{T,H}=1\,$TeV at the LHC, Higgs bosons are mainly produced in gluon-gluon fusion (ggF), followed by production in association with a heavy vector boson ($VH$). Vector boson fusion (VBF) and $t\tb H$ production remain subleading \cite{Buckley:2021gfw}. Extrapolating to higher $p_{T,H}$, $VH$ production becomes the dominant production channel, followed by VBF and $t\tb H$ production. Considering only $t\tb H$ production, the tails of the $p_{T,H}$ distribution receive contributions from regions of the phase space where the Higgs becomes quasi-collinear to one of the top-quarks, leading to large logarithms involving the ratio of the masses and the hard scale $p_{T,H}$. These logarithms can grow arbitrarily large at arbitrarily high transverse momenta, potentially spoiling the perturbative convergence. Although this has not been demonstrated explicitly for $t\overline{t}H$ production, such large logarithms from quasi-collinear singularities are known to spoil perturbative convergence in the case of heavy-quark production \cite{Cacciari:1998it}. For heavy-quark production, it is well known how to resum such logarithms to all orders in perturbation theory using the framework of perturbative fragmentation functions \cite{Mele:1990cw}, recovering a convergent result. In essence, the framework involves starting from a calculation where all masses are set to zero, followed by a systematic massification. The extension of this framework to Higgs boson production was first presented in Ref.~\cite{Braaten:2015ppa}.

However, this framework is only accurate up to corrections in the ratio of the masses and the hard scale of the process, i.e.~up to power corrections of $\mathcal{O}(m_{t}/p_{T,H},m_{H}/p_{T,H})$. The impact of this approximation was studied for $t\overline{t}H$ production at leading order in Ref.~\cite{Braaten:2015ppa}, but only for the quark-anti-quark annihilation production channel. The approximation was found to yield accurate results above energies of about 600 GeV. However, if this framework is to be used for realistic cross section calculations, it needs to be verified to be accurate for the full proton-proton cross section, i.e.~including all possible initial states. Furthermore, since state-of-the-art predictions for this process go well beyond leading order, the accuracy of the perturbative fragmentation-function formalism must be verified beyond leading order if it is to be relied on for accurate resummed predictions in the future. The goal of this paper is to verify under which conditions the perturbative fragmentation-function formalism can reliably describe $t\overline{t}H$ production through NLO in QCD.\footnote{The perturbative fragmentation-function formalism can easily be applied to other Higgs boson production channels, such as $VH$ production. However, the first fragmentation contributions would only arise at NLO. Instead, we focus on $t\tb H$ production where the factorisation theorem can be applied already at LO due to the 3-particle final state.}

The paper is organised as follows. In Section~\ref{sec:theory}, we present the formulation of the perturbative fragmentation formalism for $t\overline{t}H$ production. In Section~\ref{sec:MasslessttH}, we discuss how to evaluate cross sections with $m_t=m_H=0$. We present our results in Section~\ref{sec:Results} and conclude in Section~\ref{sec:Conclusions}. Two appendices describe the massless LO calculations in more detail and sketch our implementation of a subtraction scheme for scalar particles at NLO.

\section{The perturbative fragmentation formalism}
\label{sec:theory}
Fragmentation functions were initially introduced to describe hadron production at high $p_T$ \cite{Berman:1971xz}. They are the final state equivalent of parton distribution functions (PDFs) and can be used to capture the non-perturbative QCD dynamics of hadronisation to light hadrons. For this, they are usually fitted to data in the same fashion as PDFs (see e.g.\ Refs.~\cite{Moffat:2021dji,AbdulKhalek:2022laj,Gao:2025hlm}).
A fragmentation function $D_{i\rightarrow h}(z)$ can be interpreted as the number density of a hadron $h$ being produced from a parton $i$ with longitudinal momentum fraction $z$. The single-inclusive production of a hadron $h$ at the LHC is described by the cross section
\begin{align}
    \D\sigma_{pp\to h+X}=\sum_{a,b,i}\int_0^1&\D x_1\D x_2\D z\,f_a(x_1,\mu_F)f_b(x_2,\mu_F)\notag\\&\times\D\sigma_{ab\to i+X'}\left(x_1,x_2,p_i=\frac{p_h}{z},\mu_F,\mu_{Fr}\right)D_{i\rightarrow h}(z,\mu_{Fr})+\mathcal{O}\left(\frac{\Lambda_{\text{QCD}}}{p_{T,h}}\right),\label{eq:FactorisationNP}
\end{align}
where $f_a$ is the proton PDF for the parton $a$, and the sum over $a$, $b$, and $i$ runs over all massless partons. $\D\sigma_{ab\to i+X'}$ is the single-inclusive partonic cross section for producing parton $i$ from the collision of partons $a$ and $b$. Eq.~\eqref{eq:FactorisationNP} holds up to power corrections in the non-perturbative hadronisation scale $\Lambda_{\text{QCD}}$.

The identification of a particular collinear parton violates the inclusivity requirement of the Kinoshita-Lee-Nauenberg (KLN) theorem \cite{Kinoshita:1962ur,Lee:1964is}. If dimensional regularisation with $d = 4-2\epsilon$ dimensions is used to regulate divergences in the cross section, then this violation leads to the non-cancellation of poles in $\epsilon$. However, as for PDFs, the structure of these poles is process-independent and can be predicted to all orders in perturbation theory, allowing them to be removed via a collinear renormalisation of the fragmentation functions. This renormalisation introduces a new arbitrary scale, the fragmentation scale $\mu_{Fr}$. The dependence of the fragmentation functions on $\mu_{Fr}$ is governed by the time-like DGLAP evolution equations. As usual, the finite terms included in the renormalisation determine the renormalisation scheme. The most common scheme, modified minimal subtraction ($\msb$), will be used here.

In the early 90s, it was noticed that the fragmentation formalism could also be applied to the collinear radiation effects appearing in the production of a heavy quark \cite{Mele:1990cw}. Here, we discuss its application to top-quarks, as this will be the relevant heavy quark throughout this work. There are three distinct sources of finite-mass effects:
\begin{enumerate}
    \item collinear effects involving a final-state top-quark;
    \item collinear effects involving initial-state partons;
    \item ultraviolet effects in virtual corrections.
\end{enumerate}
The factorisation of the partonic cross section can be written as
\begin{multline}
\label{eq:topQuarkFactorisation}
    \D\sigma_{ab\rightarrow t+X}(m_t,p_t;\alpha_S^{(5)})
    =
    \sum_{a',b',i}f_{a\rightarrow a'}\otimes f_{b\rightarrow b'}\otimes\D\sigma_{a'b'\rightarrow i+X'}(m_t=0;\alpha_S^{(6)})\otimes D_{i\rightarrow t} +\mathcal{O}\left(\frac{m_t^2}{p_{T,t}^2}\right),
\end{multline}
where we defined the convolutions
\begin{align}
    \label{eq:convFF}
    \D\sigma_{ab\rightarrow i+X}\otimes D_{i\rightarrow t} =& \;\int_0^1\D z\,\D\sigma_{ab\rightarrow i+X}\left(p_i=\frac{p_t}{z}\right)D_{i\rightarrow t}(z)\;,\\[0.2cm]\notag\label{eq:convPDF}
    f_{a\rightarrow a'}\otimes f_{b\rightarrow b'}\otimes\D\sigma_{a'b'\rightarrow i+X}=&\\\int_0^1\D x_1\D x_2\,&f_{a\rightarrow a'}(x_1)f_{b\rightarrow b'}(x_2)\D\sigma_{a'b'\rightarrow i+X}(p_{a'}=x_1p_a,p_{b'}=x_2p_b)\;,
\end{align}
and we omitted the dependences on the factorisation and fragmentation scales $\mu_F$ and $\mu_{Fr}$ for brevity. The sum is over all partons, including the (anti-)top-quark. The first source of finite-mass effects is related to collinear effects in the final state. Such effects reside purely within the fragmentation functions $D_{i\rightarrow t}$. Due to the KLN theorem, there can be no sensitivity to emissions collinear to particles which are not explicitly identified, i.e.~particles other than the two initial-state partons and the identified top-quark in the final state. As a result, no fragmentation functions have to be included in the factorisation theorem for the other final-state partons. Furthermore, since the factorisation theorem is used to describe the single-inclusive production of top-quarks at high transverse momenta, the emission of soft top-quark pairs does not alter the observed final state. Such soft emissions and their corresponding finite-mass effects thus also vanish by the KLN theorem.

The second source of leading mass dependence arises due to collinear effects involving the initial state, such as an anti-top-quark becoming collinear to an initial-state parton. This is taken into account by including perturbative heavy-quark matching coefficients $f_{a\rightarrow a'}$ \cite{Aivazis:1993pi}. For example, an effective top PDF is generated at $\mathcal{O}(\alpha_S)$ from the matching coefficient
\begin{equation}
    f_{g\rightarrow t}(x;\mu_F) = \frac{\alpha_ST_F}{2\pi}(x^2+(1-x)^2)\ln{\frac{\mu_F^2}{m_t^2}}+\mathcal{O}(\alpha_S^2)
\end{equation}
as follows
\begin{align}\notag
    \label{eq:topPDF}
    f_t(x;\mu_F)=&\;\int_0^1\D y\D z\,\delta(x-yz)f_g(z;\mu_F)f_{g\rightarrow t}(y;\mu_F)\\[0.2cm]
    =&\;\int_0^1\frac{\D y}{y}\,f_g(x/y,\mu_F)f_{g\rightarrow t}(y;\mu_F)\;.
\end{align}
Some other relevant matching coefficients are
\begin{align}
    f_{g\rightarrow g}(x;\mu_F)={}&\delta(1-x)-\frac{\alpha_S}{2\pi}\frac{2}{3}T_F\ln{\frac{\mu_F^2}{m_t^2}}\delta(1-x)+\mathcal{O}(\alpha_S^2)\;,\\[0.2cm]
    f_{q\rightarrow q}(x;\mu_F)={}&f_{\qb\rightarrow\qb}(x;\mu_F)=\delta(1-x)+\mathcal{O}(\alpha_S^2)\;,
\end{align}
with all remaining coefficients being $\mathcal{O}(\alpha_S^2)$ \cite{Buza:1995ie,Buza:1996wv,Bierenbaum:2009zt}. The introduction of the heavy-quark matching coefficients can be viewed as changing the PDF flavour scheme from the 5- to the 6-flavour scheme.

The final leading top mass effects arise from ultraviolet effects in virtual corrections involving top-quarks. They are taken into account by a scheme change in $\alpha_S$ from the 5- to the 6-flavour scheme. The two schemes are related by the decoupling relation \cite{Weinberg:1980wa,Ovrut:1980dg}
\begin{equation}
    \alpha_S^{(6)}(\mu_R)=\alpha_S^{(5)}(\mu_R)\left(1+\frac{\alpha_S^{(5)}(\mu_R)}{3\pi}T_F\ln{\frac{\mu_R^2}{m_t^2}}+\mathcal{O}(\alpha_S^2)\right),
\end{equation}
where we have made the dependence on the renormalisation scale $\mu_R$ explicit. Notably, for $t\tb$ production at NLO in the $gg$ channel, the contributions arising from the scheme changes in the gluon PDF and $\alpha_S$ cancel for $\mu_R=\mu_F$.

All dependence on $m_t$ is thus factorised into the fragmentation functions, the PDFs and $\alpha_S$. These objects all satisfy their own renormalisation group equations, enabling the resummation of large logarithms of $m_t$. However, the factorisation theorem Eq.~\eqref{eq:topQuarkFactorisation} only holds up to power corrections in the top-quark mass, limiting the accuracy of resummed predictions at energies not much larger than the top-quark mass. These power corrections can be taken into account at fixed order using prescriptions such as ACOT \cite{Aivazis:1993kh,Aivazis:1993pi}, S-ACOT \cite{Kramer:2000hn} or FONLL \cite{Cacciari:1998it}.

The application of the fragmentation formalism to single-inclusive Higgs production is straightforward:\footnote{A similar idea to using perturbative fragmentation to approximate the $t\overline{t}H$ production cross section at NLO was first put forward in Ref.~\cite{Dawson:1997im}. In that work, the authors determined the LO top-to-Higgs fragmentation function in the $m_H=0$ limit for a specific choice of $\mu_{Fr}$ and approximated the NLO fragmentation function by including virtual and soft gluon corrections. However, the authors only considered contributions to the $t\overline{t}H$ production cross section involving a fragmenting top or anti-top quark, which corresponds to picking the terms with $i=t,\tb$ in the sum in Eq.~\eqref{eq:FactorisationH}, missing leading power contributions already at LO.}
\begin{align}\label{eq:FactorisationH}
    \notag
    \D\sigma_{ab\rightarrow H+X}=\sum_{a',b',i}&f_{a\rightarrow a'}\otimes f_{b\rightarrow b'}\otimes\D\sigma_{a'b'\rightarrow i+X'}\left(m_H=m_t=0,p_i=\frac{p_H}{z}\right)\notag\\&\otimes D_{i\rightarrow H}(z)+\mathcal{O}\left(\frac{m_H^2}{p_{T,H}^2},\frac{m_t^2}{p_{T,H}^2}\right),
\end{align}
where the sum over particles now includes all massless partons, the (anti-)top-quark and the Higgs boson. The change in flavour-number scheme in the couplings is not explicitly shown for brevity, but must be accounted for. The top Yukawa is renormalised in the on-shell scheme whenever the top-quark is massive and in the $\msb$ scheme if the top-quark is massless. We account for this scheme change using \cite{Braaten:1980yq,Tarrach:1980up}
\begin{align}
    y_t^{(\msb)}=y_t^{(\text{OS})}\bigg(1-\frac{3\alpha_S}{4\pi}C_F\bigg(\ln{\frac{\mu_R^2}{m_t^2}}+\frac{4}{3}\bigg)+\mathcal{O}(y_t^2,\alpha_S^2)\bigg)\;.
\end{align}
Note that all particles must be taken as massless on the RHS of Eq.~\eqref{eq:FactorisationH}. We will only consider interactions involving the strong coupling and the top-quark Yukawa coupling. If e.g.~interactions involving electroweak gauge bosons are considered as well, then these bosons must also be taken as massless on the RHS of Eq.~\eqref{eq:FactorisationH}, and they must be included in the sum over particles. Note that when considering differential distributions in $p_{T,H}$, the convolution takes the usual form
\begin{equation}
    \frac{\sigma_{ab\rightarrow i + X}}{\D p_{T,H}}\otimes D_{i\rightarrow H}=\int_0^1\frac{\D z}{z}\,\frac{\D\sigma_{ab\rightarrow i + X}}{\D p_{T,i}}\left(p_i=\frac{p_H}{z}\right)D_{i\rightarrow H}(z)\;,
\end{equation}
with an additional factor $1/z$ due to $\D p_{T,H}=z\,\D p_{T,i}$.

At LO in QCD, only the terms from $i=t,\tb,H$ contribute to Eq.~\eqref{eq:FactorisationH}. The respective fragmentation functions are given by \cite{Braaten:2015ppa,Brancaccio:2021gcz}
\begin{align}
    D_{H\rightarrow H}(z) &= \delta(1-z)+\mathcal{O}(y_t^2)\;,\\[0.2cm]
    D_{\tb\rightarrow H}(z)&=D_{t\rightarrow H}(z) =\notag\\&\frac{y_t^2}{16\pi^2}z
    \bigg[ - \ln\bigg(\frac{m_t^2}{\mu^2}\bigg)
      - \ln\left(z^2 + 4\xi^2(1-z)\right)
      + 4(1-\xi^2)\frac{1-z}{z^2+4\xi^2(1-z)}\bigg]+\mathcal{O}(y_t^2\alpha_s)\;,
\end{align}
with the top Yukawa $y_t=\frac{m_t}{v}$ and the ratio $\xi=\frac{m_H}{2m_t}$. $D_{g\rightarrow H}$, which is non-zero only beyond LO, and $D_{t\to H}$ are both known through NLO \cite{Brancaccio:2021gcz}. The fragmentation functions involving other quark flavours would start contributing at NNLO. Note that the leading logarithm is proportional to $y_t^2$: resummation at leading-logarithmic accuracy would thus include all terms of the form $y_t^{2n}\ln^n(\mu^2/p^2_{T,H})$.

We will be interested in single-inclusive Higgs production in association with a top-quark pair specifically.\footnote{More specifically, we will require at least one top-quark pair in the final state. The discussion for the case of requiring exactly one top-quark pair to be present is different, but analogous considerations apply.} However, requiring the presence of a real top-quark pair in the final state significantly complicates the factorisation theorem. While logarithmically enhanced contributions are only present in cases where the Higgs boson is produced in association with a top-quark pair through NLO, this ceases to be the case starting at NNLO. Starting at that order, diagrams where the Higgs boson is produced via a virtual top-quark loop can also contribute to the cross section, as long as a top-quark pair is produced elsewhere in the event via gluon splitting. This will introduce additional collinear effects, not related to the Higgs boson at all: if a gluon far away from the Higgs boson in phase space splits into a quasi-collinear top-quark pair, this will generate logarithmic effects. If the presence of a top-quark pair was not required, then these logarithmic effects would cancel against those coming from a virtual top-quark loop on the same gluon leg. However, since a real top-quark pair must be present for a contribution to be included in the cross section, this virtual correction does not contribute and the logarithmic effects associated with the gluon splitting to a top-quark pair do not cancel.

One might be tempted to modify the factorisation formula by adding a second collinear final-state factor accounting for such collinear effects unrelated to the Higgs boson, yielding a cross section of the form
\begin{align}\label{eq:FactorisationH2}
&\sum_{a',b',i}f_{a\rightarrow a'+X_1}\otimes f_{b\rightarrow b'+X_2}\otimes\D\sigma_{a'b'\rightarrow i+X_\sigma}\otimes D_{i\rightarrow H+X_D}\bigg\rvert_{t\overline{t}\in\bigcup_kX_k}\notag\\&+\sum_{a',b',i,j}f_{a\rightarrow a'+X_1}\otimes f_{b\rightarrow b'+X_2}\otimes\D\sigma_{a'b'\rightarrow ij+X_\sigma}\otimes D_{i\rightarrow H+X_D}\bigg\rvert_{t\overline{t}\notin\bigcup_kX_k}\times F_{j\rightarrow t\overline{t}+X_F}\;,
\end{align}
where each object now comes in two variants: one where a top-quark pair is produced and one where no top-quarks are produced, which is implicitly marked using the notation $X_i$ for particles produced in association with the one explicitly identified. E.g., the usual PDF matching condition $f_{g\to g}$ receives contributions from diagrams where a real top-quark pair is produced, as well as from diagrams where the top-quark only contributes in virtual loops. Usually, these two cases do not need to be distinguished, since one is not interested in the heavy quarks produced this way. But now this distinction is necessary. The summation over all possible sets $X_i$ is implicit in Eq.~\eqref{eq:FactorisationH2}. The cross section then receives two types of contributions: one where the top-quark pair is produced somewhere in the usual factorisation formula, corresponding to the first term in Eq.~\eqref{eq:FactorisationH2}, and one where it is not produced in the hard process or any of the explicitly identified collinear splittings, but via another collinear splitting of another final-state particle $j$, which corresponds to the second term in Eq.~\eqref{eq:FactorisationH2}. This term is multiplied by the fragmentation fraction $F_{j\to t\overline{t}}$, which is the total probability for a particle $j$ to produce a top-quark pair through collinear splittings. Note that this is a number, not a function, since we do not measure the top-quarks' momenta. At NLO, the gluon fragmentation fraction is given by the integral over the perturbative fragmentation function (PFF) $D_{g\to t}$. However, this is not true in general, since the fragmentation functions are defined as single-inclusive quantities, i.e.~a splitting producing two top-quarks contributes twice to the PFF. The fragmentation fractions would only count such states once.

However, the naive generalisation shown in Eq.~\eqref{eq:FactorisationH2} is incorrect, as it leads to double counting. E.g., starting from a final state containing two well-separated gluons, labelled $g_1$ and $g_2$, and a Higgs boson, there are three pairs of collinear splittings at relative $\mathcal{O}(\alpha_S^2)$ with mass effects which could produce at least one top-quark pair:
\begin{align}
&g_1\to t\overline{t},&&g_2\overset{\text{top-loop}}{\to}g_2;\notag\\
&g_1\overset{\text{top-loop}}{\to}g_1,&&g_2\to t\overline{t};\notag\\
&g_1\to t\overline{t},&&g_2\to t\overline{t}.\notag\\
\end{align}
Despite the fact that the third splitting produces two top-quark pairs, each splitting should be counted only once. By using Eq.~\eqref{eq:FactorisationH2}, one is again implicitly using the KLN theorem to sum over all other splittings. In the above scenario, this would double-count the third splitting.

The formula which correctly accounts for all collinear effects would be
\begin{align}\label{eq:FactorisationH3}
&\sum_n\sum_{a',b',i,j_1,...,j_n}f_{a\rightarrow a'+X_1}\otimes f_{b\rightarrow b'+X_2}\otimes\D\sigma_{a'b'\rightarrow ij_1...j_n}\otimes D_{i\rightarrow H+X_D}\prod_{m=1}^n F_{j_m\rightarrow X_{j_m}}\bigg\rvert_{t\overline{t}\in\bigcup_kX_k}\;,
\end{align}
where again the sum over all possible sets of associated particles $X_i$ is implicit.

Although already significantly more complex than Eq.~\eqref{eq:FactorisationH}, this form only captures collinear effects. In addition, there are also soft effects, resulting from the emission of soft top-quark pairs. As mentioned above, soft effects cancel in single-inclusive cross sections due to the KLN theorem. However, in the present case, the emission of a soft top-quark pair would alter the observable state, again spoiling the cancellation. Indeed, the PDF matching conditions, the PFF and the fragmentation fractions in Eq.~\eqref{eq:FactorisationH3} all contain soft singularities coming from such soft emissions, leading to higher poles in $\epsilon$ than expected from functions capturing collinear effects. Furthermore, splittings to a top-quark pair described via the fragmentation fractions also correspond to soft emissions if the splitting particle is itself soft. In addition to all these soft-collinear emissions, wide-angle soft emissions must be accounted for as well. This can be done using the $S$ function formalism introduced in Ref.~\cite{Generet:2025gdy}. Schematically, one introduces a set of operators $\mathbf{S}_{X_S}$, which account for the mass effects related to the soft emission of the set of particles $X_S$. Since the soft-collinear modes are already accounted for in the collinear functions, they must be subtracted from $\mathbf{S}_{X_S}$. Denoting this properly subtracted $S$ function as $\mathbf{\tilde{S}}_{X_S}$, the final factorisation formula is then
\begin{align}\label{eq:FactorisationH4}
\D\sigma_{ab\rightarrow t\overline{t}H+X}(p_H)=&\sum_n\sum_{a',b',i,j_1,...,j_n}f_{a\rightarrow a'+X_1}\otimes f_{b\rightarrow b'+X_2}\otimes\big(\mathbf{\tilde{S}}_{X_s}\cdot\D\sigma_{a'b'\rightarrow ij_1...j_n}\big)\notag\\&\otimes D_{i\rightarrow H+X_D}\prod_{m=1}^n F_{j_m\rightarrow X_{j_m}}\bigg\rvert_{t\overline{t}\in\bigcup_kX_k}+\mathcal{O}\left(\frac{m_H^2}{p_{T,H}^2},\frac{m_t^2}{p_{T,H}^2}\right)\;,
\end{align}
where $\cdot$ indicates that $\mathbf{\tilde{S}}_{X_S}$ is a colour operator to be inserted in the product of matrix elements implicit in $\D\sigma$. Instead of subtracting the soft-collinear modes from the $S$ functions, they could also have been subtracted from the collinear functions, as is typically done in SCET \cite{Manohar:2006nz}. However, since the present case would also require corresponding subtractions in $\D\sigma$, it seems more transparent to perform these subtractions in the $S$ functions. Note that the presence of soft-collinear logarithms means that the behaviour of the perturbative series is double-logarithmic: at sufficiently high orders, each additional order in perturbation theory raises the highest power of mass logarithms by two.

One more comment about this factorisation formula is in order. As usual, this factorisation assumes QCD colour coherence, i.e.~it assumes that soft and collinear emissions can be treated independently. However, it is known \cite{Catani:2011st,Forshaw:2012bi,Schwartz:2017nmr,Henn:2024qjq,Guan:2024hlf} that QCD colour coherence breaks down at sufficiently high orders in perturbation theory in many hadron collider processes. Assuming this to be the case for $t\overline{t}H$ production as well, the above factorisation formula is, in fact, still incorrect and the real factorisation formula is even less transparent.

From this discussion, it should be apparent that the factorisation theorem is significantly more complicated once additional constraints are put on the final state. However, we will only be interested in applying the factorisation theorem through NLO, i.e.~$\mathcal{O}(y_t^2\alpha_s^3)$. In that case, Higgs bosons can only be produced in association with a top-quark pair via diagrams where the Higgs boson couples to the top-quark line. This means that there are no soft-top-quark contributions and no collinear mass effects resulting from final-state splittings unrelated to the Higgs boson. This means that the much simpler factorisation formula for single-inclusive production, Eq.~\eqref{eq:FactorisationH}, can be used.

Up to NLO, the factorisation theorem for the partonic cross section is thus given by
\begin{align}
    \notag
    \label{eq:Masterformula}
    \D\sigma_{ab\rightarrow t\tb H}(m_t,m_H)\bigg\rvert_{\alpha_S^{(5)},y_t^{(\text{OS})}}
    =\; \bigg[&f_{a\to a'}\otimes f_{b\to b'}\otimes\D\sigma_{a'b'\rightarrow H+X}(m_t=m_H=0)\notag\\
     &+ 2\, f_{a\to a'}\otimes f_{b\to b'}\otimes\D\sigma_{a'b'\rightarrow t+X}(m_t=0)\otimes D_{t\rightarrow H}(m_t,m_H)\notag\\
     &+\,\D\sigma_{ab\rightarrow g+X}\otimes D_{g\rightarrow H}(m_t,m_H)\bigg]\bigg\rvert_{\alpha_S^{(6)},y_t^{(\msb)}}\;,
\end{align}
where we used $D_{t\to H}=D_{\overline{t}\to H}$\footnote{We will only consider cross sections which are symmetric in rapidity, so that mirrored asymmetric initial states can be accounted for by a factor of 2.} and the fact that only $D_{t\to H}$ and $D_{g\to H}$ are non-zero through this order. The change in flavour-number scheme is now explicitly shown. For clarity and brevity of notation, we omit the dependence on the momenta and the scales. Instead, we make the dependence on the masses explicit. The first term corresponds to the fragmentation of a Higgs boson to a Higgs boson, where the PFF can be omitted, since $D_{H\to H}=\delta(1-z)$. In this term, we do not include contributions where the Higgs boson couples to virtual top-quark loops, since in that case no real top-quark pair can be produced through NLO. We will refer to the first term as the direct contribution and to the other terms as the fragmentation contributions. In the third term, we used that $D_{g\rightarrow H}$ is only non-zero starting at NLO, so that there can be no contributions coming from the PDF matching conditions in this term through NLO.

At LO, two partonic channels contribute: $q\overline{q}$ and $gg$. For these channels, and at LO, Eq.~\eqref{eq:Masterformula} takes the form
\begin{align}
\label{eq:MasterformulaLO}
    \D\sigma^{\lo}_{q\overline{q}\rightarrow t\tb H}(m_t,m_H)
    \approx{}& \D\sigma^{\lo}_{q\overline{q}\rightarrow t\tb H}(m_t=m_H=0)
     + 2\,\D\sigma^{\lo}_{q\overline{q}\rightarrow t\tb}(m_t=0)\otimes
     D^{\lo}_{t\rightarrow H}(m_t,m_H),\\
     \D\sigma^{\lo}_{gg\rightarrow t\tb H}(m_t,m_H)
    \approx{}& \D\sigma^{\lo}_{gg\rightarrow t\tb H}(m_t=m_H=0)+4f^{(1)}_{g\to t}\otimes\D\sigma^{\lo}_{tg\rightarrow t H}(m_t=m_H=0)\notag\\
     &+ 2\,\D\sigma^{\lo}_{gg\rightarrow t\tb}(m_t=0)\otimes
     D^{\lo}_{t\rightarrow H}(m_t,m_H),
\end{align}
where the superscript $(n)$ denotes the N$^n$LO QCD coefficient of a quantity. The factor of 4 in front of the contribution proportional to $f_{g\to t}$ takes into account that either of the top-quarks can become collinear to either of the initial-state gluons.

At NLO, three initial-state channels contribute: $q\overline{q}$, $gg$ and $qg$ (and $\overline{q}g$, which can be treated identically to $qg$). Their explicit expressions can be trivially derived from Eq.~\eqref{eq:Masterformula} by inserting the perturbative expansions for the PDF matching conditions, PFFs, partonic cross sections and threshold matching conditions for the couplings and keeping only the terms proportional to $y_t^2\alpha_s^3$.

We will label cross sections computed according to the factorisation theorem as Zero-Mass Top-Quark (ZMTQ) cross sections. These will be contrasted with cross sections computed according to the hybrid prescription, which accounts for the leading power corrections in the top-quark mass and is described in the following subsection.

\subsection{Hybrid prescription}
\label{sec:theoryHybrid}
The factorisation formula is only accurate up to power corrections in $m_t$ and $m_H$. In order to more clearly separate the impact of power corrections in different masses, one can compute cross sections using a hybrid prescription \cite{Braaten:2015ppa}, including the leading power corrections in the top-quark mass.

Schematically, cross sections are computed according to the hybrid prescription as follows:
\begin{align}\label{eq:Hybrid}
\D\sigma_{pp\to t\overline{t}{H}}&(m_t,m_H) = \D\sigma_{pp\to t\overline{t}{H}}(m_t,0)\notag\\&{}+\D\sigma^\text{ZMTQ}_{pp\to t\overline{t}{H}}(m_t,m_H)-\D\sigma^\text{ZMTQ}_{pp\to t\overline{t}{H}}(m_t,0)+\mathcal{O}\bigg(\frac{m_H^2}{p_{T,H}^2},\frac{m_H^2m_t^2}{p_{T,H}^4}\bigg)\;,
\end{align}
where $\D\sigma^\text{ZMTQ}$ is computed according to the factorisation theorem. There are two complementary ways of viewing this formula. The first is that, starting from the factorisation theorem, one adds the leading power corrections in $m_t$ by adding the difference between the result retaining the exact $m_t$-dependence at $m_H=0$ and its factorised approximation. The second viewpoint is that, starting from the cross section with $m_t\neq0$ and $m_H=0$, one first subtracts the leading-power mass effects in $m_t$ using the factorisation theorem, recovering, up to power corrections, the massless $t\overline{t}H$ production cross sections. Then one inserts this approximation of the massless cross section in the factorisation formula, adding the leading-power mass effects in $m_t$ and $m_H$ simultaneously.

Trivially, Eq.~\eqref{eq:Hybrid} reproduces the exact $m_t$-dependence in the $m_H\to0$ limit. Since all quantities involved are smooth functions of $m_H$ in that limit, the hybrid prescription indeed includes all leading power corrections in $m_t$, i.e.~within the hybrid prescription, there can be no power corrections that scale as pure powers of $m_t/p_{T,H}$. All power corrections in $m_t$ are necessarily suppressed by $m_H/p_{T,H}$.

The explicit formula for the cross section according to the hybrid prescription can again be trivially obtained by inserting the perturbative expansions of all relevant ingredients and truncating through $y_t^2\alpha_s^3$. In practice, all contributions in the factorised cross sections where the Higgs boson is produced in the hard scattering process cancel in the difference of factorised cross sections in Eq.~\eqref{eq:Hybrid}, since those terms contain no dependence on $m_H$. All the other terms, i.e.~the terms proportional to $D_{t\to H}$ or $D_{g\to H}$, trivially simplify to terms proportional to differences of fragmentation functions, since these are the only source of $m_H$-dependence in the factorised cross section. For a general single-inclusive Higgs boson production cross section, and at LO in $y_t$ but to all orders in QCD, the hybrid prescription can thus also be written as
\begin{align}\label{eq:FactorisationHybrid}
    \notag
    \D\sigma_{ab\rightarrow H+X}(m_t,m_H)={}&\D\sigma_{ab\rightarrow H+X}(m_t,0)+\sum_{a',b',i}f_{a\rightarrow a'}\otimes f_{b\rightarrow b'}\otimes\D\sigma_{a'b'\rightarrow i+X'}\left(m_t=0,p_i=\frac{p_H}{z}\right)\notag\\&{}\otimes \bigg[D_{i\rightarrow H}(z,m_t,m_H)-D_{i\rightarrow H}(z,m_t,0)\bigg]+\mathcal{O}\left(\frac{m_H^2}{p_{T,H}^2},\frac{m_H^2m_t^2}{p_{T,H}^4}\right),
\end{align}
where the contribution from $i=H$ trivially vanishes and can thus be omitted.

\section{Massless $t\tb H$ calculation}
\label{sec:MasslessttH}
The direct contributions of the ZMTQ factorisation theorem~\eqref{eq:Masterformula} include the processes $q\qb\rightarrow t\tb H$ and $gg\rightarrow t\tb H$ evaluated at $m_t=m_H=0$. In addition, the processes $\qbm g\rightarrow t\tb H\qbm$ and $tg\rightarrow tHg$ are required at LO for massless tops and Higgs. In these cases, collinear divergences are no longer regulated by the masses but appear as poles in the dimensional regularisation parameter $\epsilon$. Notably, these divergences (which correspond to NLO singularity patterns) appear already at LO due to the 3-particle final state. At LO (NLO), these singularities can usually be handled by running the LO (NLO) calculation as an NLO (NNLO) calculation using some subtraction scheme. Because most subtraction schemes are built to handle only QCD radiation, the presence of the massless scalar Higgs boson is an issue. In this section, we describe different methods to deal with this problem, separately for the LO and the NLO calculation of massless $t\tb H$ production.

\subsection{The LO case}
\label{sec:MasslessttHLO}
Only the $q\qb$ and the $gg$ channels contribute to the LO calculation of $t\tb H$ production. For this setup, the 3-particle final state $t\tb H$ can approach one (soft-) collinear limit at a time. The possible singularity structures are therefore NLO-like and can be captured by an NLO subtraction scheme that allows for scalar particles. We included such a scheme involving scalar particles in \verb|STRIPPER|, the \verb|C++| implementation of the sector-improved residue subtraction scheme \cite{Czakon:2014oma,Czakon:2010td,Czakon:2011ve,Czakon:2019tmo}. Within the \verb|STRIPPER| formalism, such an implementation is rather straightforward since soft, collinear and soft-collinear radiation are treated independently and the integration over the unresolved phase space of the integrated subtraction terms is done numerically. The implementation is sketched briefly in Appendix~\ref{app:ScalarSubtraction}.

While sufficient for the tests of the full factorisation theorem at LO, the general subtraction scheme suffers from missed-binning effects. These effects appear because the momenta of the Higgs in the evaluation of the matrix element and the subtraction term coincide only in the strict collinear limit. Close to a collinear limit and a bin edge, it can happen that the regular term and the subtraction term end up in different bins of the spectrum and thus no subtraction is effectively done in a region of phase space where it would be numerically necessary. It can be shown that this problem can lead to non-converging Monte-Carlo errors if sharp bin edges are used. Typically, one can mitigate these effects by distributing each weight to two bins if they are close to a bin edge. This procedure, called bin smearing, leads to a converging Monte-Carlo error but can distort the distribution slightly (usually on the sub-percent level). In practice, a rather large smearing parameter is required for the scalar subtraction scheme in order to achieve acceptably small errors. This leads to distortions that are for example too strong for a high-precision test of the method presented in the next section for the NLO calculation. We also determine the LO fully-massless cross sections semi-analytically. The calculation in the $q\qb$ channel can be done easily up to a final integration over the Higgs rapidity (and the convolution with the PDFs) which is done numerically. For the $gg$ channel, customised subtraction terms are defined which allow for numerical integration without missed-binning effects. The calculations for both channels are presented in Appendix~\ref{app:masslessttH}.

\subsection{The NLO case}
\label{sec:MasslessttHNLO}
At NLO, final states such as $t\tb Hg$ appear. Hence, the singularity structure corresponds to an NNLO calculation. In principle, it would be possible to extend the \verb|STRIPPER| library by scalar sectors at NNLO. However, we did not pursue this route. Instead, we rely on the factorisation theorem to approximate the fully-massless direct contribution. Since the factorisation theorem holds for any top-quark mass up to power corrections, we can pick a small top-quark mass $\mts$ and $m_H=0$ to minimise power corrections. The massless direct contribution is then obtained by rearranging Eq.~\eqref{eq:Masterformula} and replacing $m_t$ with $\mts$ and setting $m_H=0$. We choose the numerical value $\mts=10\,$GeV to find a balance between small power corrections and numerical stability. This balance is necessary because a smaller mass requires larger cancellations between the direct and fragmentation contributions and therefore a higher numerical precision. Due to the complex logarithmic structure of the real corrections at NLO in the direct contribution, a high numerical precision is difficult to achieve as is apparent from the discussion in Sec.~\ref{sec:Results}.

\begin{figure}
    \centering
    \includegraphics[width=0.49\linewidth]{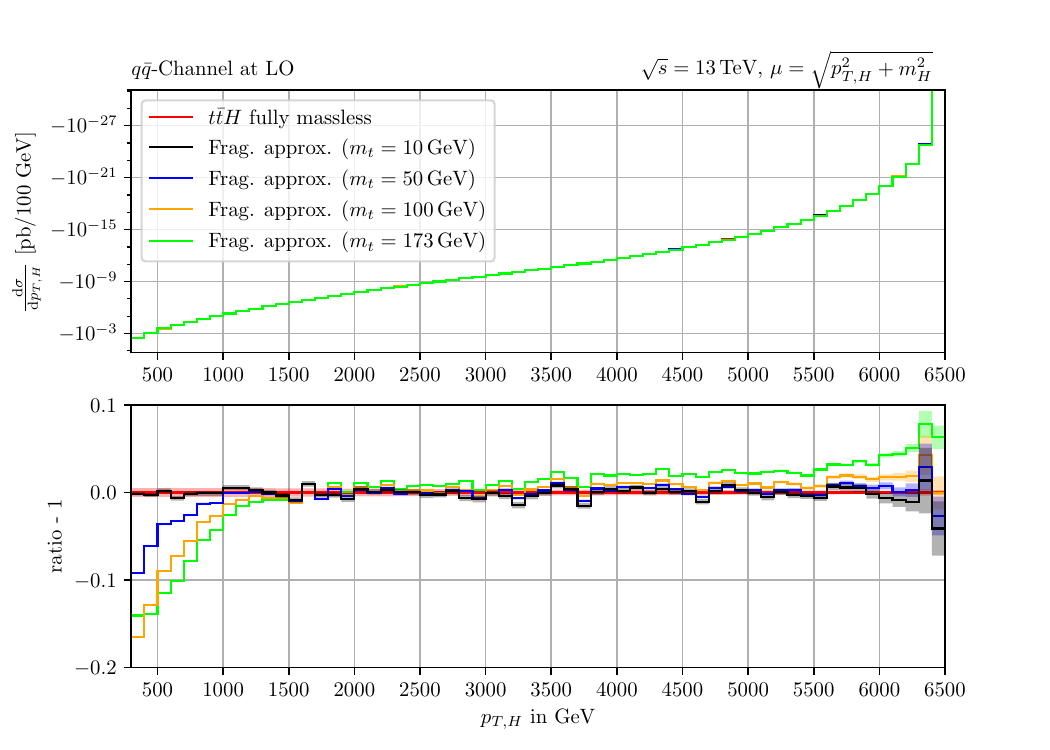}
    \includegraphics[width=0.49\linewidth]{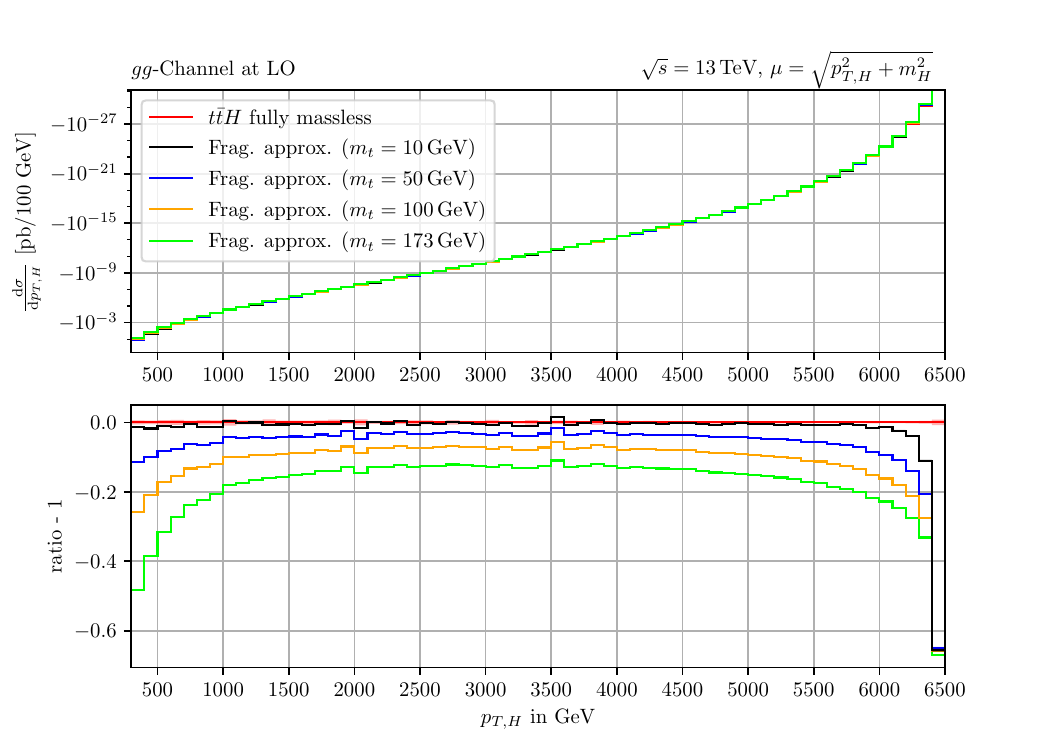}
    \includegraphics[width=0.49\linewidth]{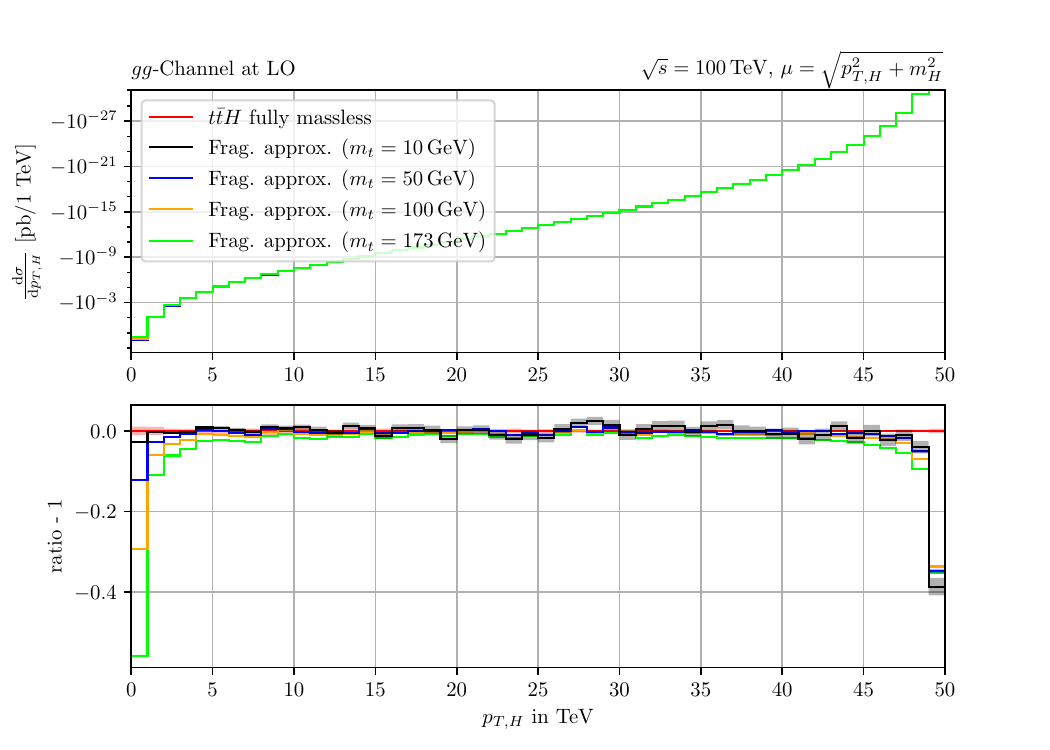}
    
    \caption{Verification that at LO the massless calculation is approximated well using a small top-quark mass in the factorisation theorem. Different values of the top-quark mass in the approximation are compared to the semi-analytic massless calculation (red) and ratios are taken with respect to this calculation. Shown are the $q\qb$ channel at $E_{\text{cms}}=13\,$TeV (top left), the $gg$ channel at $E_{\text{cms}}=13\,$TeV (top right) and the $gg$ channel at $E_{\text{cms}}=100\,$TeV (bottom). The calculations are performed using the setup of Section~\ref{sec:Results} and $m_H=0$ everywhere except within the dynamical scale where we use $m_H=125\,$GeV. The shaded bands indicate the Monte Carlo integration error.}
    \label{fig:anaPlots}
\end{figure}

Writing all terms explicitly, in the $q\qb$ channel, the massless direct contribution is given by
\begin{align}\notag
    \D\sigma^{\nlo}_{q\qb\rightarrow t\tb H}(0,0)=&\;\D\sigma^{\nlo}_{q\qb\rightarrow t\tb H}\left(\mts,0\right)
    -2\,\D\sigma^{\nlo}_{q\qb\rightarrow t\tb}\left(\mts\right)\otimes D^{\lo}_{t\rightarrow H}\left(\mts,0\right)\\[0.1cm]\notag
    &-2\,\D\sigma^{\lo}_{q\qb\rightarrow t\tb}\left(\mts\right)\otimes D^{\nlo}_{t\rightarrow H}\left(\mts,0\right)\\[0.1cm]
    &-2\,\D\sigma^{\lo}_{q\qb\rightarrow gg}\otimes D^{\nlo}_{g\rightarrow H}\left(\mts,0\right)-2\frac{\alpha_S}{3\pi}T_F\ln{\frac{\mu_R^2}{(\mts)^2}}\,\D\sigma^{\lo}_{q\qb\rightarrow t\tb H}(0,0)\\[0.1cm]\notag
    &+\frac{3\alpha_S}{2\pi}C_F\bigg(\ln{\frac{\mu_R^2}{(\mts)^2}}+\frac{4}{3}\bigg)\,\D\sigma^{\lo}_{q\qb\rightarrow t\tb H}(0,0)\;,
\end{align}
where the final two terms make the scheme changes of $\alpha_S$ and $y_t$ explicit. The LO direct contribution could be evaluated with one of the methods described in the previous section. For consistency, we express it via the factorisation theorem as well. It can be approximated by
\begin{equation}
    \D\sigma^{\lo}_{q\qb\rightarrow t\tb H}(0,0)=\D\sigma^{\lo}_{q\qb\rightarrow t\tb H}\left(\mts,0\right)
    -2\,\D\sigma^{\lo}_{q\qb\rightarrow t\tb}\left(\mts\right)\otimes D^{\lo}_{t\rightarrow H}\left(\mts,0\right).
\end{equation}
In the $gg$ channel, additional terms appear due to initial-state collinear configurations. The direct contribution is given by\footnote{Since we use $\mu_R=\mu_F$ here, the contributions coming from the $\alpha_S$ decoupling constant and the gluon-to-gluon PDF threshold matching condition cancel and are omitted.}
\begin{align}\notag
    \D\sigma^{\nlo}_{gg\rightarrow t\tb H}(0,0)=&\;\D\sigma^{\nlo}_{gg\rightarrow t\tb H}\left(\mts,0\right)
    -2\,\D\sigma^{\nlo}_{gg\rightarrow t\tb}\left(\mts\right)\otimes D^{\lo}_{t\rightarrow H}\left(\mts,0\right)\\[0.1cm]\notag
    &-2\,\D\sigma^{\lo}_{gg\rightarrow t\tb}\left(\mts\right)\otimes D^{\nlo}_{t\rightarrow H}\left(\mts,0\right)\\[0.1cm]
    &-2\,\D\sigma^{\lo}_{gg\rightarrow gg}\otimes D^{\nlo}_{g\rightarrow H}\left(\mts,0\right)\\[0.1cm]\notag
    &- 4f^{\lo}_{g\rightarrow t}\left(\mts\right)\otimes\D\sigma^{\nlo}_{tg\rightarrow tH}(0,0)\\[0.1cm]\notag
    &- 4f^{\nlo}_{g\rightarrow t}\left(\mts\right)\otimes\D\sigma^{\lo}_{tg\rightarrow tH}(0,0)\\[0.1cm]\notag
    &- 2f^{\lo}_{g\rightarrow t}\left(\mts\right)f^{\lo}_{g\rightarrow\tb}\left(\mts\right)\otimes\D\sigma^{\lo}_{t\tb\rightarrow Hg}(0,0)\\[0.1cm]\notag
    &+ \frac{3\alpha_S}{2\pi}C_F\bigg(\ln{\frac{\mu_R^2}{(\mts)^2}}+\frac{4}{3}\bigg)\,\D\sigma^{\lo}_{gg\rightarrow t\tb H}(0,0)\;.
\end{align}
The LO direct contribution is given by
\begin{align}\notag
    \D\sigma^{\lo}_{gg\rightarrow t\tb H}(0,0)=&\;\D\sigma^{\lo}_{gg\rightarrow t\tb H}\left(\mts,0\right)
    -4f^{\lo}_{g\rightarrow t}\left(\mts\right)\otimes\D\sigma^{\lo}_{tg\rightarrow tH}(0,0)\\[0.1cm]
    &-2\,\D\sigma^{\lo}_{gg\rightarrow t\tb}\left(\mts\right)\otimes D^{\lo}_{t\rightarrow H}\left(\mts,0\right).
\end{align}
The direct contribution of the $qg$ channels becomes
\begin{align}
    \notag
    \D\sigma^{\lo}_{qg\rightarrow t\tb Hq}(0,0)=&
    \;\D\sigma^{\lo}_{qg\rightarrow t\tb Hq}\left(\mts,0\right)
    -2\,\D\sigma^{\lo}_{qg\rightarrow t\tb q}\left(\mts\right)\otimes
    D^{\lo}_{t\rightarrow H}\left(\mts,0\right)\\[0.1cm]\notag&
    -\D\sigma^{\lo}_{qg\rightarrow qg}\otimes
    D^{\nlo}_{g\rightarrow H}\left(\mts,0\right)\\[0.1cm]
    &- 2f^{\nlo}_{q\rightarrow t}\left(\mts\right)\otimes\D\sigma^{\lo}_{tg\rightarrow tH}(0,0)\\[0.1cm]\notag
    &-2f^{\lo}_{g\rightarrow t}\left(\mts\right)\otimes\D\sigma^{\lo}_{tq\rightarrow tHq}(0,0)\;.
\end{align}
This approach can be verified at LO by comparison to the explicit calculation described in Sec.~\ref{sec:MasslessttHLO} and Appendix~\ref{app:masslessttH}. In Fig.~\ref{fig:anaPlots}, the semi-analytic massless calculations at LO and the approximation using the factorisation theorem are shown for varying values of the small top-quark mass $\mts$. It is clear that a value $\mts=10\,$GeV approximates the fully-massless calculation very well over the bulk range of the spectrum.

\section{Results}
\label{sec:Results}
The goal of this work is to quantify the quality of the factorisation theorem for $t\tb H$ production. Therefore, we compare LO and NLO calculations with $m_H=125\,$GeV and $m_t=173\,$GeV (called the fully-massive calculation) to the approximation given by the factorisation theorem in either prescription. We do not apply DGLAP evolution to the fragmentation functions, so if the factorisation theorem were exact, the two calculations should yield identical results. Their differences thus arise due to the power corrections neglected in the factorisation theorem. We work with a center-of-mass (cms) energy of 13 TeV for both the ZMTQ and hybrid prescriptions and additionally with $E_{\text{cms}}=100\,$TeV for the ZMTQ prescription. We use \verb|NNPDF4.0| PDFs at NLO accuracy (even for LO calculations) \cite{NNPDF:2021njg}, included via the \verb|LHAPDF| interface \cite{Buckley:2014ana}, and dynamical scales $\mu_R^2=\mu_F^2=\mu_{Fr}^2=p_{T,H}^2+m_H^2$ where $m_H=125\,$GeV at all times. The one-loop matrix elements for $t\tb H$ production and for massless $t\tb$ production were evaluated using \verb|OpenLoops| \cite{Buccioni:2019sur}. All calculations were performed using \verb|STRIPPER|.
We only show Monte Carlo integration errors.

\subsection{ZMTQ prescription}
\label{sec:ResultsZMTQ}
\begin{figure}
    \centering
    \includegraphics[width=0.49\linewidth]{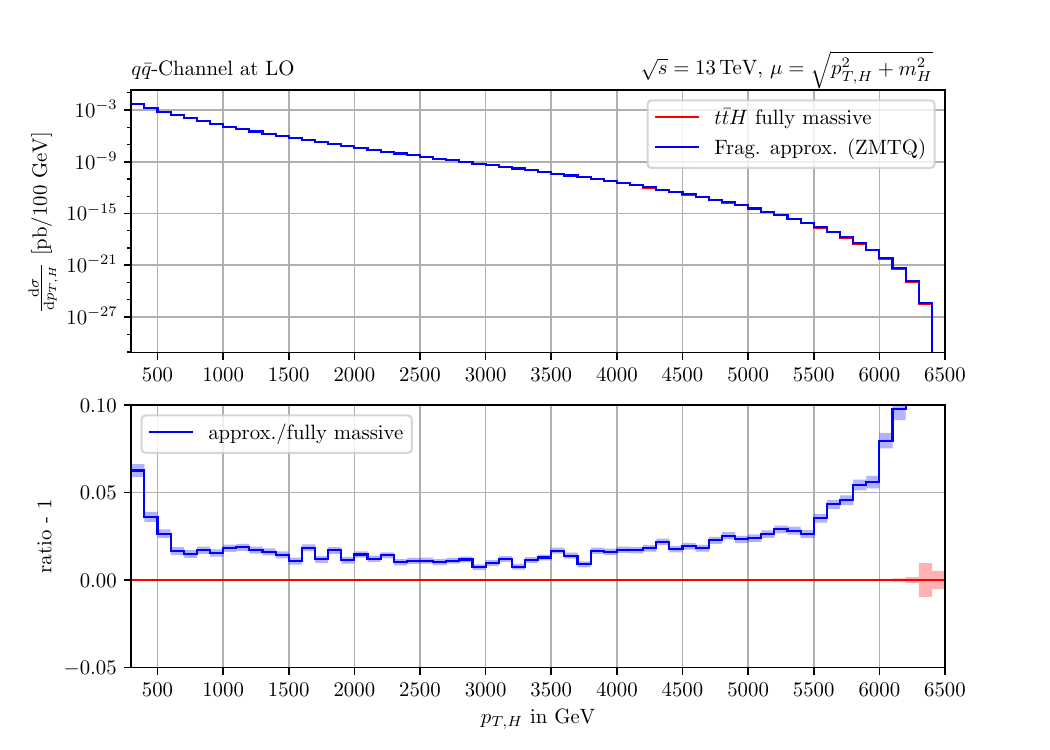}
    \includegraphics[width=0.49\linewidth]{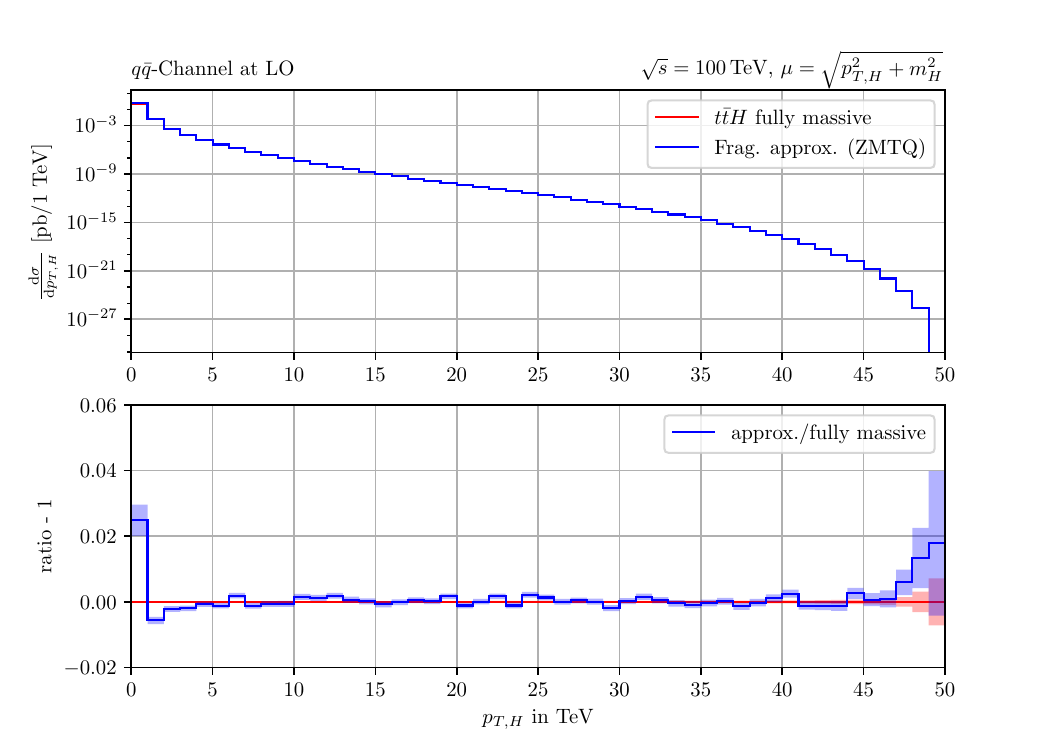}
    \caption{Comparison of the fully-massive calculation (red) and the approximation using the fragmentation formalism (blue) at LO in the $q\qb$ channel at a cms-energy of 13 TeV (left) and 100 TeV (right) in the ZMTQ prescription. Note the different scales of the ratio plots. The shaded areas show the Monte Carlo integration errors.}
    \label{fig:LOqq}
\end{figure}

\begin{figure}
    \centering
    \includegraphics[width=0.49\linewidth]{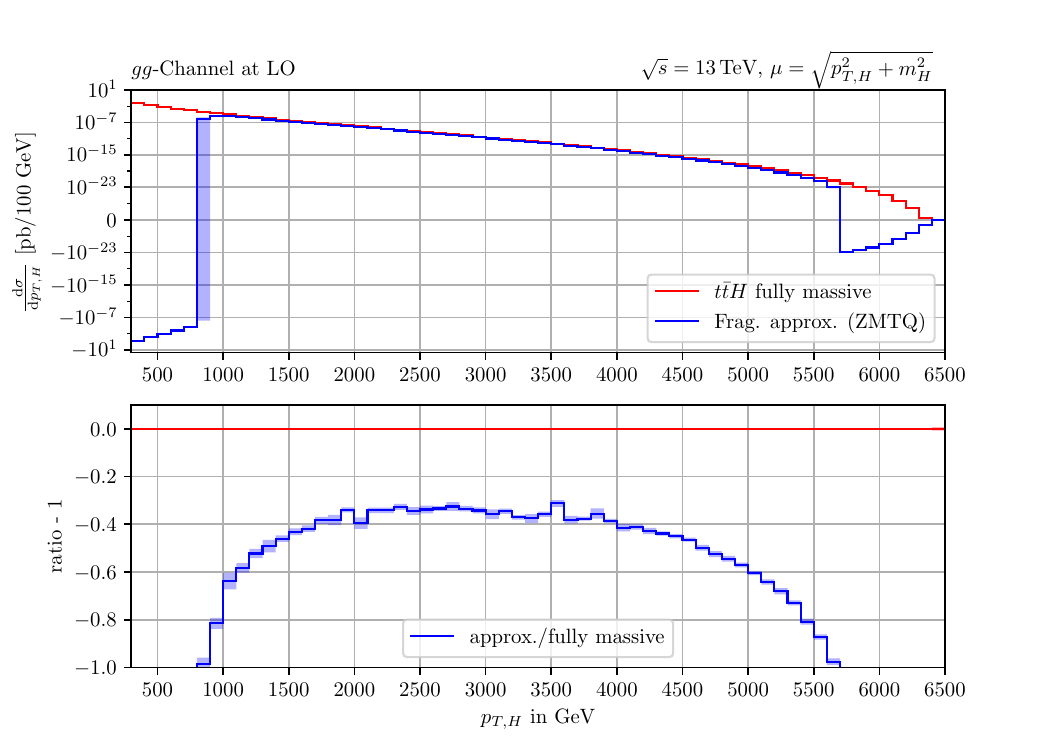}
    \includegraphics[width=0.49\linewidth]{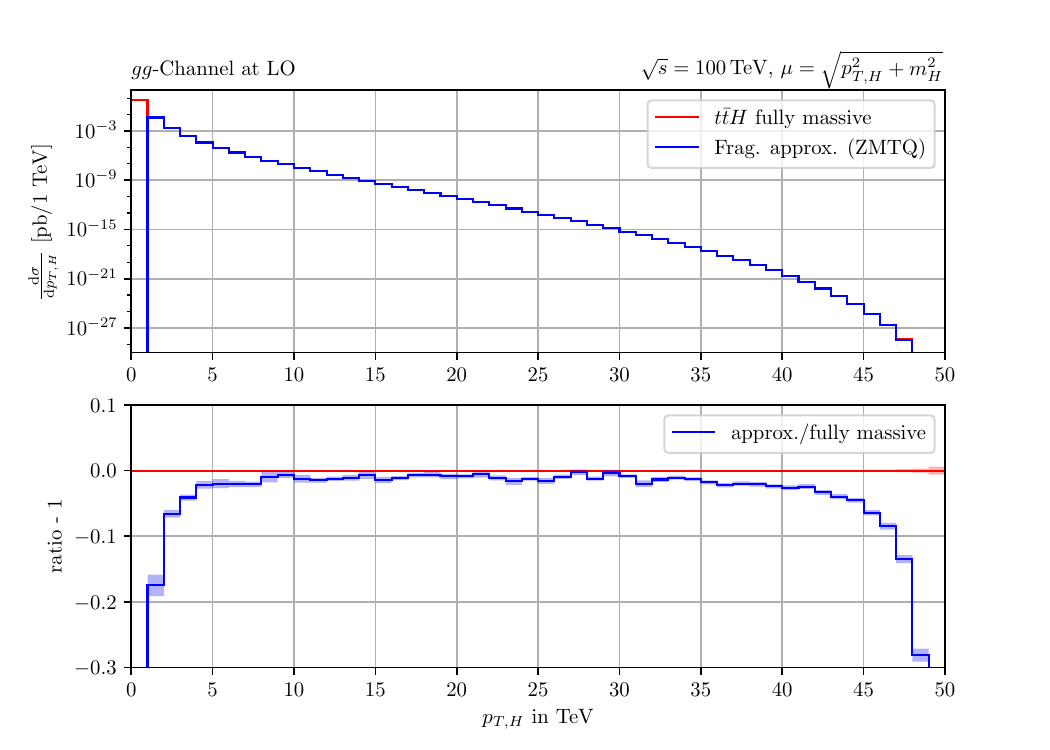}
    \caption{Same as Fig.~\ref{fig:LOqq} but in the $gg$ channel.}
    \label{fig:LOgg}
\end{figure}

The comparison between the fully-massive cross section and the cross section obtained using the factorisation theorem at LO is shown in Figs.~\ref{fig:LOqq} ($q\bar{q}$ channel) and~\ref{fig:LOgg} ($gg$ channel) for cms energies of 13 and 100 TeV. For the LO calculation, the direct calculations presented in Appendix~\ref{app:masslessttH} are used.

In the $q\qb$ channel, the two calculations agree down to 1-2 \% in the bulk part of the spectrum at 13 TeV and even better at 100 TeV cms energy. The larger discrepancies near the kinematic edge arise from a phase-space mismatch. The fragmentation calculation, in which particles are treated as massless, allows for higher transverse momenta than the calculations involving the massive particles. While this mismatch happens only very close to the kinematic edge, this edge is smeared along the spectrum by the PDFs. Therefore, these effects start playing a role already quite far away from the actual kinematic edge of the collider.

\begin{figure}
    \centering
    \raisebox{-0.5\height}{\includegraphics[width=0.75\linewidth]{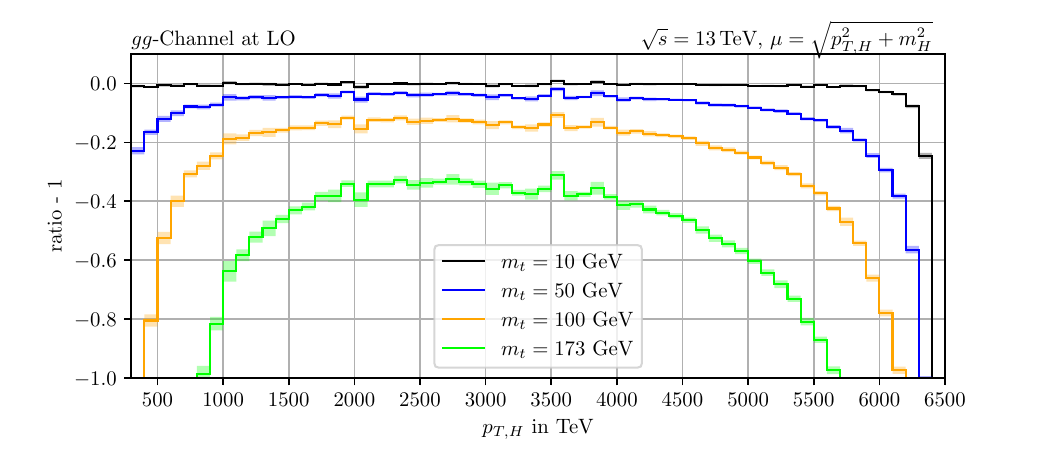}}
    \caption{Comparison of the approximation in the fragmentation formalism to the fully-massive calculation at LO in the $gg$ channel at 13 TeV for different values of the top and Higgs masses. The Higgs mass is scaled by the same factor as the top-quark mass for each calculation. The line for $m_t=173\,$GeV corresponds to the ratio plot shown on the left of Fig.~\ref{fig:LOgg}.}
    \label{fig:MassVariation}
\end{figure}

In the $gg$ channel, the agreement at 100 TeV is in the low percent range which confirms that power corrections are small at these energies. This is not the case at 13 TeV where the fragmentation calculation approximates the fully-massive calculation never better than 30 \%. To check that the reason for this is indeed power corrections, we reduced $m_t$ and $m_H$ systematically, keeping the ratio $\xi=\frac{m_H}{2m_t}$ constant, and observed a slow convergence of the fragmentation calculation towards the fully-massive one, as should be the case if the difference is purely power corrections. This comparison is shown in Fig.~\ref{fig:MassVariation}.

\begin{figure}
    \centering
    \includegraphics[width=0.49\linewidth]{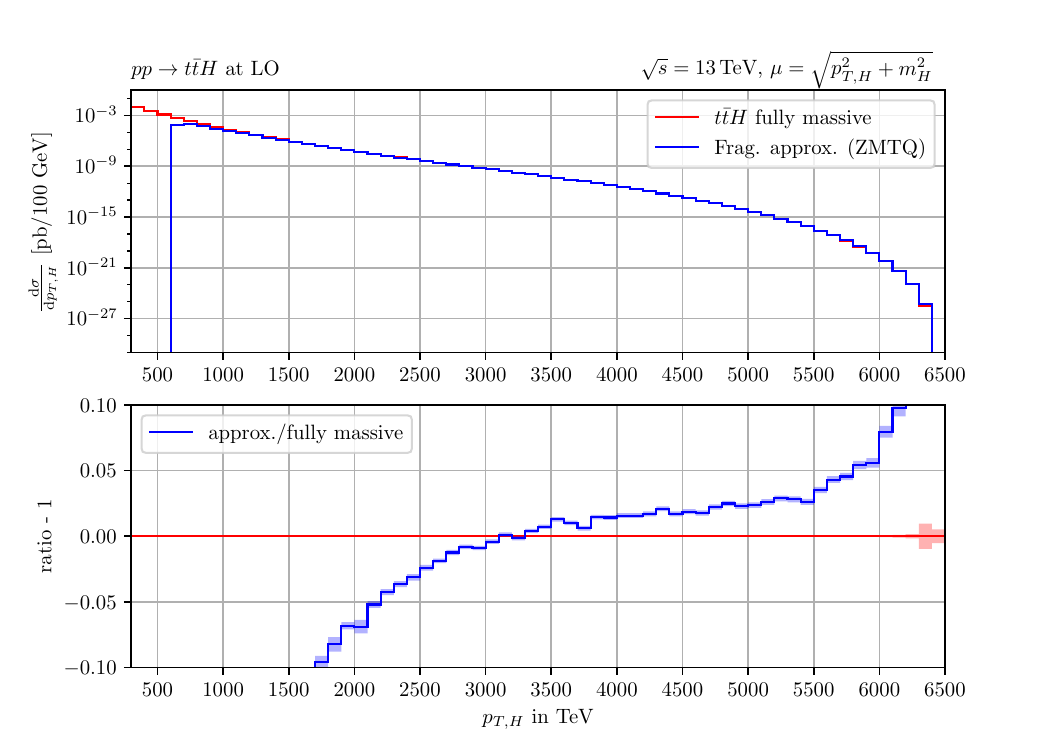}
    \includegraphics[width=0.49\linewidth]{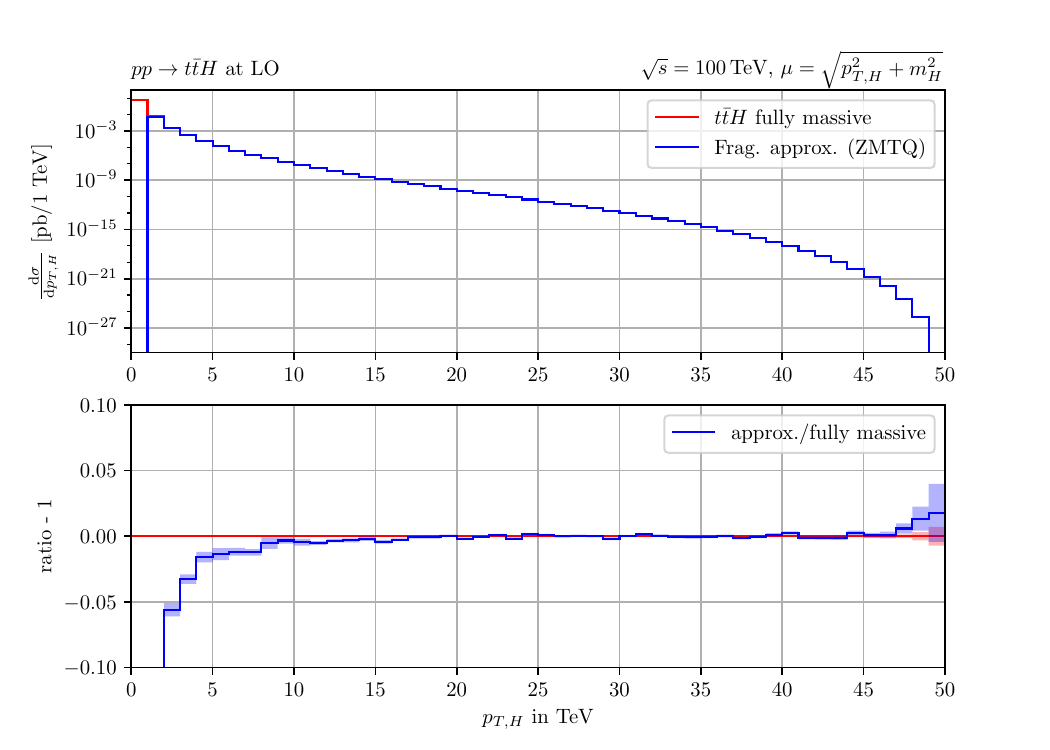}
    \caption{Same as Fig.~\ref{fig:LOqq} but with $q\qb$ and $gg$ channel combined.}
    \label{fig:LOpp}
\end{figure}
\begin{figure}
    \centering
    \includegraphics[width=0.49\linewidth]{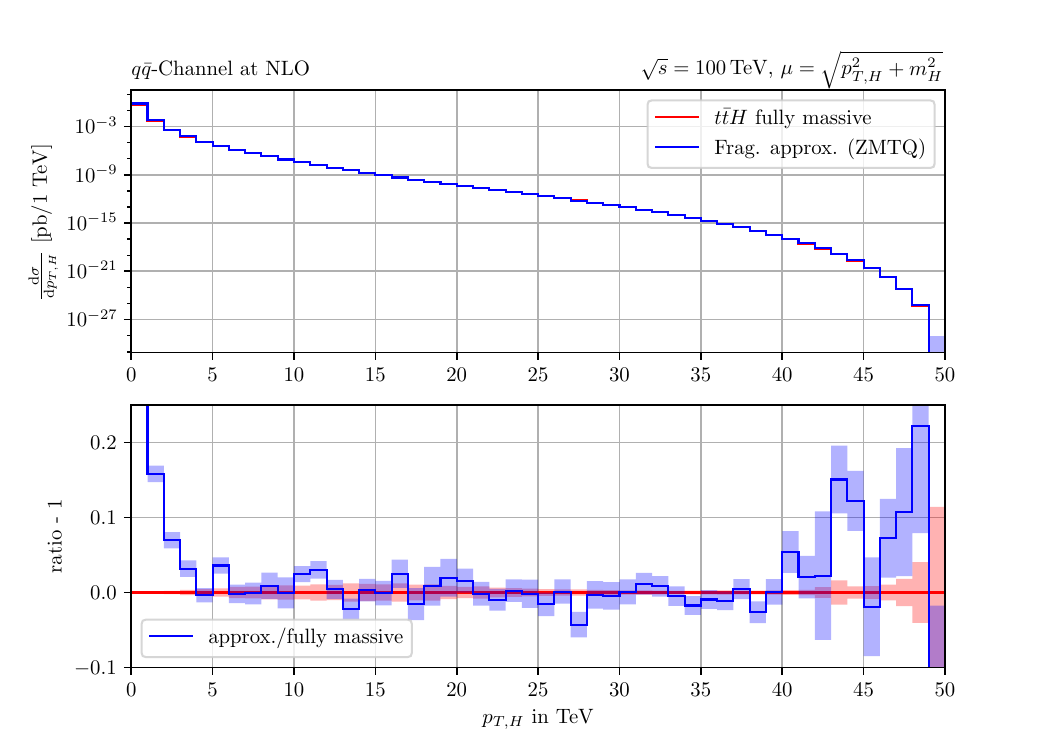}
    \includegraphics[width=0.49\linewidth]{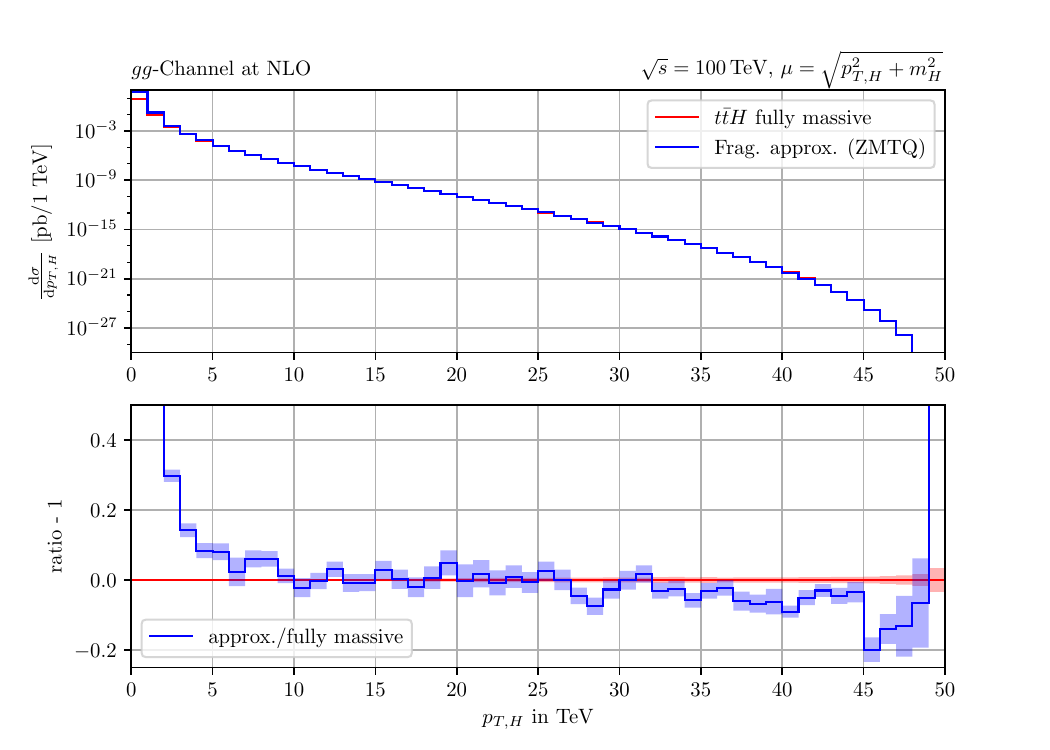}
    \caption{Comparison of the fully-massive calculation (red) and the approximation using the fragmentation formalism (blue) at NLO in the $q\qb$ channel (left) and the $gg$ channel (right) at a cms-energy of 100 TeV in the ZMTQ prescription. The shaded areas show Monte Carlo integration errors.}
    \label{fig:FCC_ZMTQ_NLO}
\end{figure}

Figure~\ref{fig:LOpp} shows the combination of the $q\qb$ and the $gg$ channels. For $13\,$TeV cms energy, the $q\qb$ channel dominates at high $p_{T,H}$ which makes the unreliable approximation in the $gg$ channel less pronounced. While the approximation is better than $5\%$ over some part of the spectrum, the ratio plot acquires a non-trivial slope due to the combination of the two channels. Going to NLO at these energies, the results would be more unreliable because the NLO corrections to the $gg$ channel are much larger than to the $q\qb$ channel, increasing the relevance of the $gg$ channel. In addition, the power corrections increase when moving to NLO, as can be seen for a cms energy of $100\,$TeV (Figs.~\ref{fig:LOqq}, \ref{fig:LOgg} and \ref{fig:FCC_ZMTQ_NLO}). Therefore, the NLO calculation was only performed at 100 TeV where the LO approximation works very well (right plot of Fig.~\ref{fig:LOpp}). Apart from the newly appearing $qg$ channels, the NLO calculation involves the gluon-to-Higgs fragmentation function and the NLO corrections to the top-to-Higgs fragmentation function. Also, we require a top PDF generated from the $\mathcal{O}(\alpha_S^2)$ heavy-quark matching coefficients which are convolved with both the gluon and the light quark PDFs. For the massless direct contributions, we use the approximation via the factorisation theorem using a small top-quark mass as described in Sec.~\ref{sec:MasslessttHNLO}. The contributions involving the real corrections $tg\rightarrow tHg$ and $tq\rightarrow tHq$ are evaluated using the scalar subtraction scheme.

\begin{figure}
    \centering
    \includegraphics[width=0.5\linewidth]{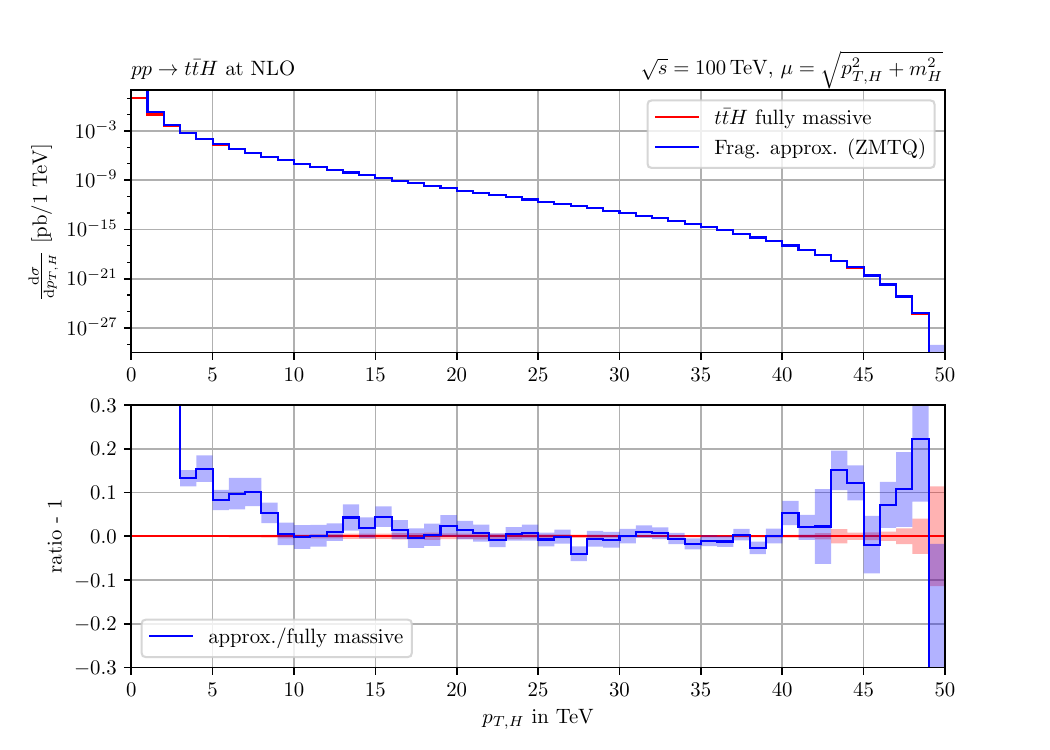}
    \caption{Same as Fig.~\ref{fig:FCC_ZMTQ_NLO} but with all channels combined.}
    \label{fig:FCC_ZMTQ_NLOpp}
\end{figure}
The $q\qb$ and $gg$ channels at NLO are shown in Fig.~\ref{fig:FCC_ZMTQ_NLO}. The fragmentation formalism is reliably able to approximate the fully-massive calculations down to a few percent. The reasons for the large Monte Carlo errors are two-fold but are connected to our method of evaluating the fully-massless cross section via the factorisation theorem. The main source of error are the real corrections with the final state $t\tb Hg$ with $m_H=0$, $m_t=10\,$GeV. While their individual relative error is small, huge cancellations between these calculations and contributions from fragmentation functions and top PDFs using the small top mass appear. On their own, the relative error of the real corrections is about 0.1 \% throughout the bulk part of the spectrum. The cancellations blow this error up by a factor 10-20.
The error itself is very hard to decrease due to the complicated logarithmic structure of the calculation. While in principle an NLO calculation, the calculation with the final state $t\bar{t}Hg$ with $m_H=0$, $m_t=10\,$GeV mimics an NNLO calculation in the sense that large logarithms of the top-quark mass appear in patterns that mimic NNLO singularity patterns. This leads to numerical instabilities which could be remedied by using massive sectors for NNLO-like phase spaces within the implementation of \verb|STRIPPER|. The inclusion of such sectors can speed up calculations by about two orders of magnitude \cite{Generet:2025gdy}. However, they were not available at the time the present work was conducted.

Fig.~\ref{fig:FCC_ZMTQ_NLOpp} shows the combination of all channels at a cms energy of $100\,$TeV. As for the individual $q\qb$ and $gg$ channels, the fully-massive $pp\rightarrow t\tb H$ calculation is approximated very well by the factorisation theorem in the bulk region of the spectrum verifying that power corrections are indeed small at these energies.

\subsection{Hybrid prescription}
\begin{figure}
    \centering
    \includegraphics[width=0.49\linewidth]{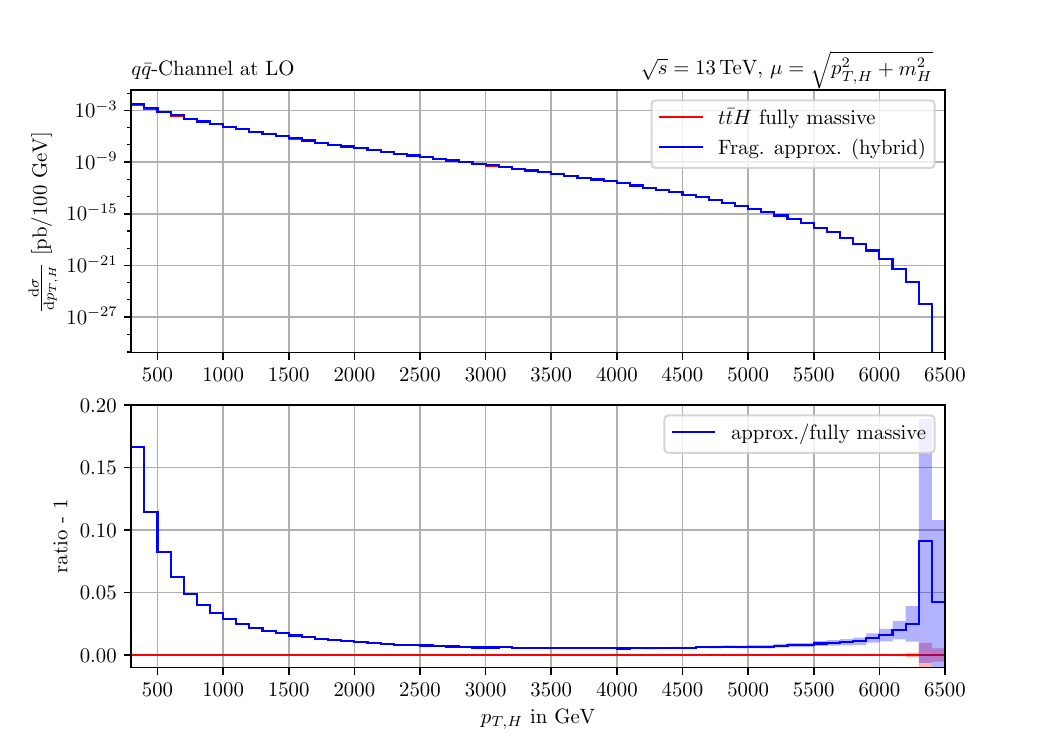}
    \includegraphics[width=0.49\linewidth]{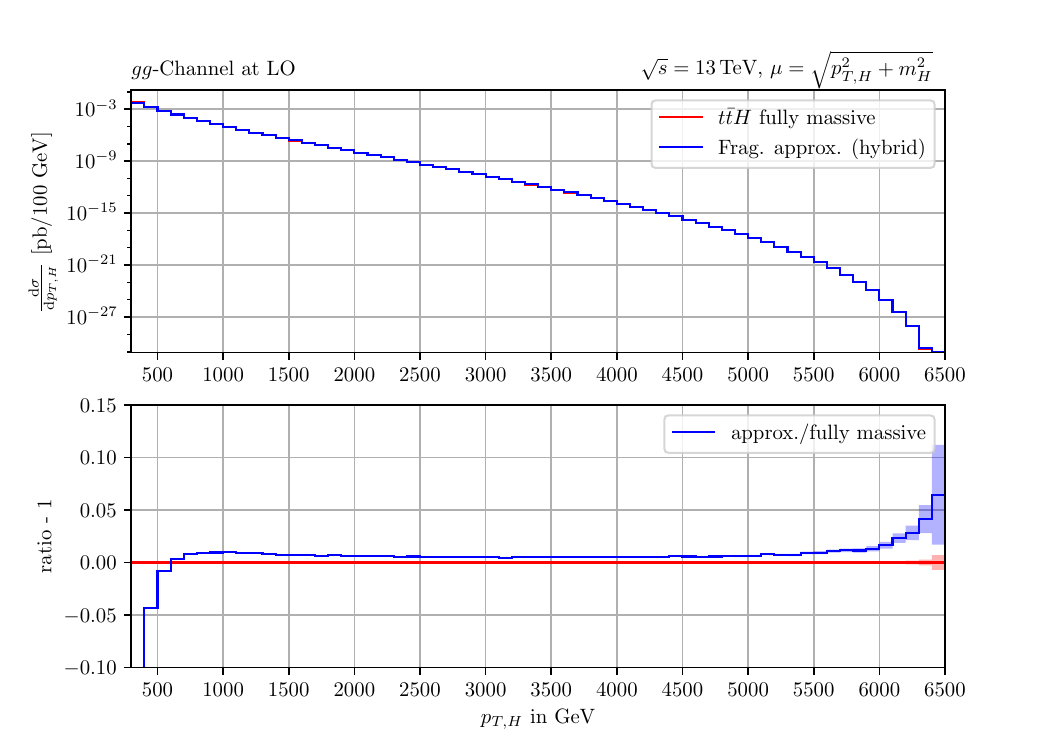}
    \caption{Comparison of the fully-massive calculation (red) and the approximation using the fragmentation formalism (blue) at LO in the $q\qb$ channel (left) and the $gg$ channel (right) at a cms-energy of 13 TeV in the hybrid prescription. The shaded areas show Monte Carlo integration errors.}
    \label{fig:LOHybridqqgg}
\end{figure}

\begin{figure}
    \centering
    \includegraphics[width=0.5\linewidth]{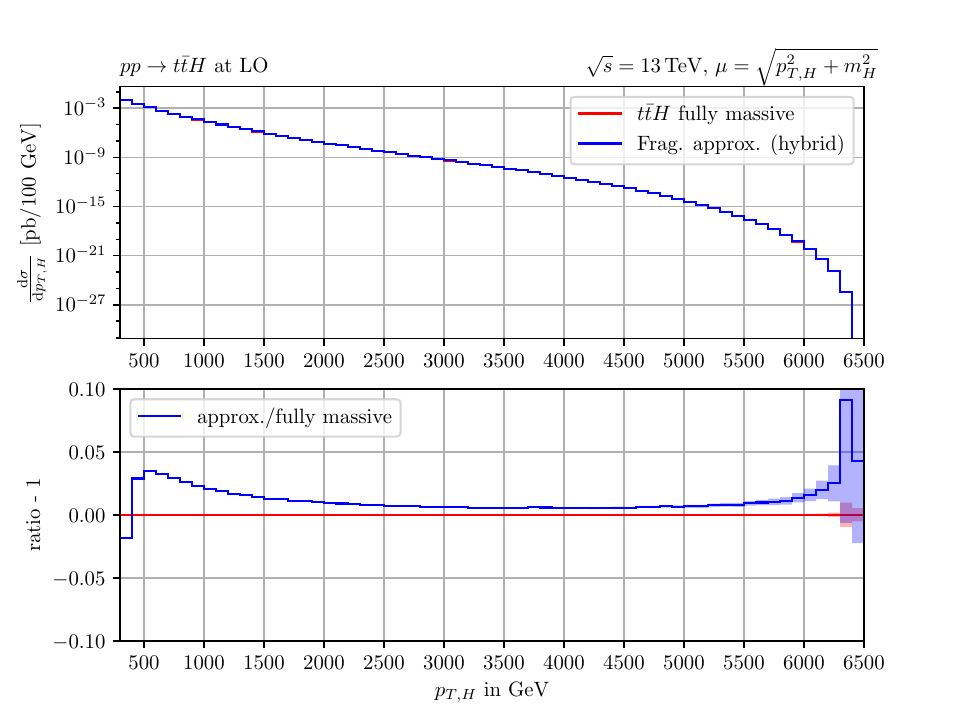}
    \caption{Same as Fig.~\ref{fig:LOHybridqqgg} but with all channels combined.}
    \label{fig:LOHybridpp}
\end{figure}

As discussed in Section~\ref{sec:theoryHybrid}, we can use a hybrid prescription to restore the exact top mass dependence for $m_H=0$ in the direct contribution. For this prescription, Fig.~\ref{fig:LOHybridqqgg} shows the LO results at a cms-energy of $13\,$TeV in the individual partonic channels and Fig.~\ref{fig:LOHybridpp} the combination of the channels. In contrast to the ZMTQ prescription, power corrections are well under control in both channels. This shows that the power corrections of the general factorisation theorem are dominated by top-quark mass effects. Also, the kinematic edge effects are less pronounced because the phase space mismatch is smaller compared to the general factorisation theorem. The numerical uncertainties increase close to the kinematic edge due to the very low sampling rate.

\begin{figure}
    \centering
    \includegraphics[width=0.49\linewidth]{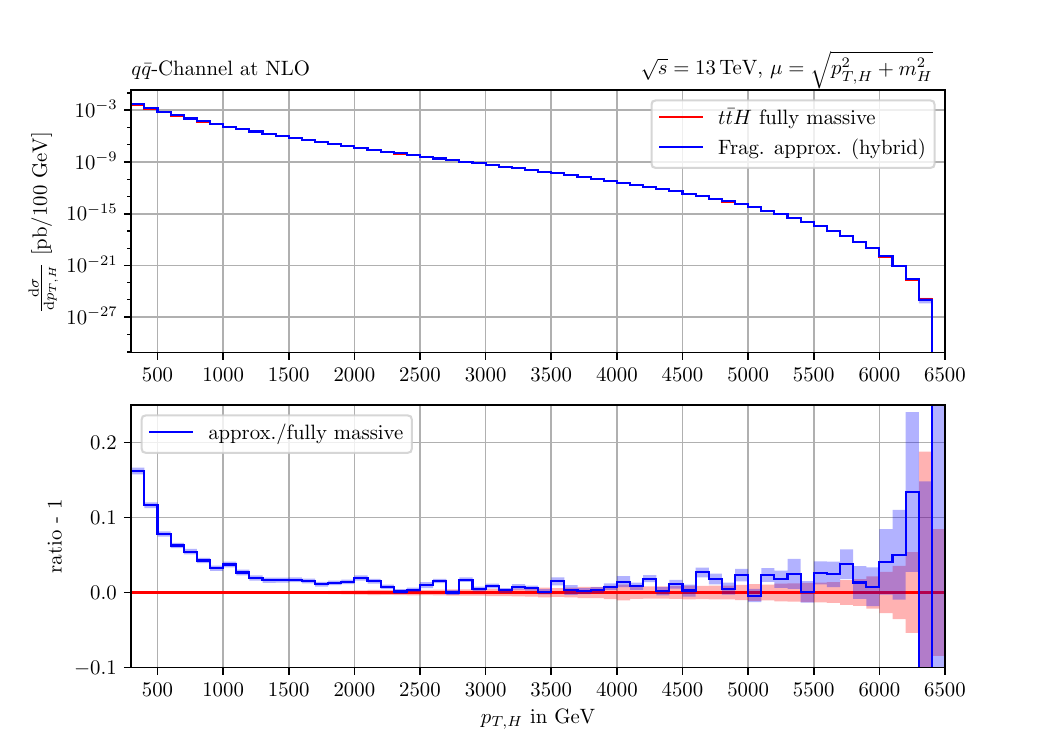}
    \includegraphics[width=0.49\linewidth]{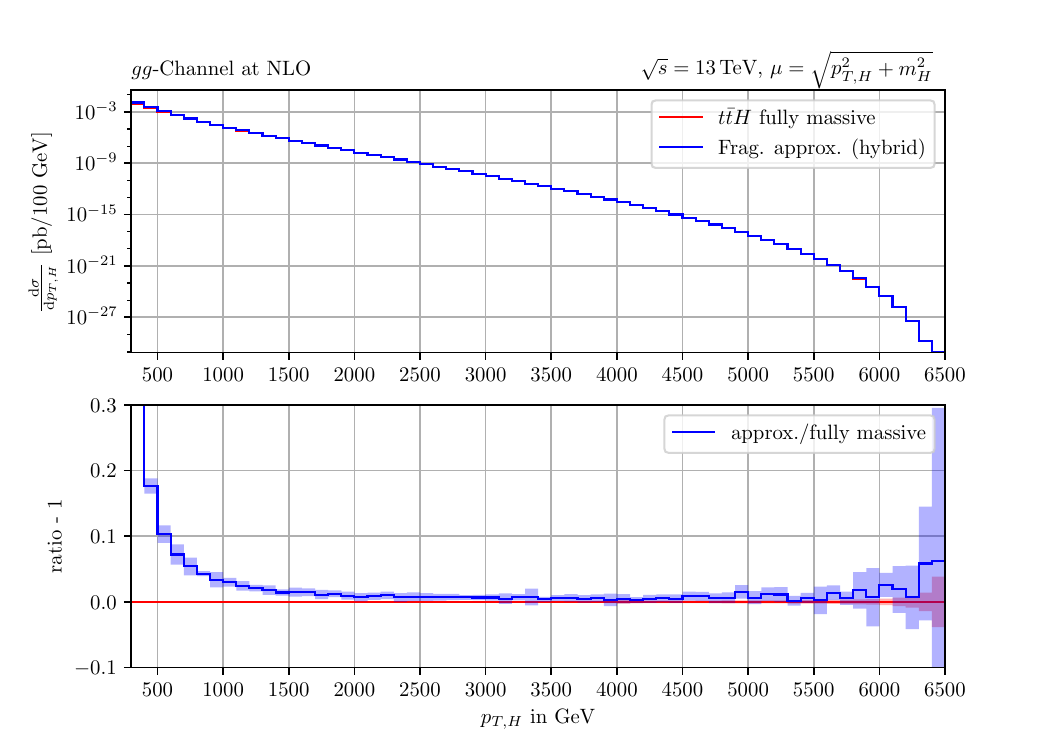}
    \caption{Comparison of the fully-massive calculation (red) and the approximation using the fragmentation formalism (blue) at NLO in the $q\qb$ channel (left) and the $gg$ channel (right) at a cms-energy of 13 TeV in the hybrid prescription. The shaded areas show Monte Carlo integration errors.}
    \label{fig:LHC_Hybrid_NLO}
\end{figure}
The NLO results of the $q\qb$ and $gg$ channel are shown in Fig.~\ref{fig:LHC_Hybrid_NLO}. It is clear that the fragmentation formalism is capable of reproducing the fully-massive calculation over a wide range of the $p_{T,H}$ spectrum in both channels down to the percent level. The NLO calculation includes the gluon-to-Higgs fragmentation function and requires a LO top PDF.

\begin{figure}
    \centering
    \includegraphics[width=0.5\linewidth]{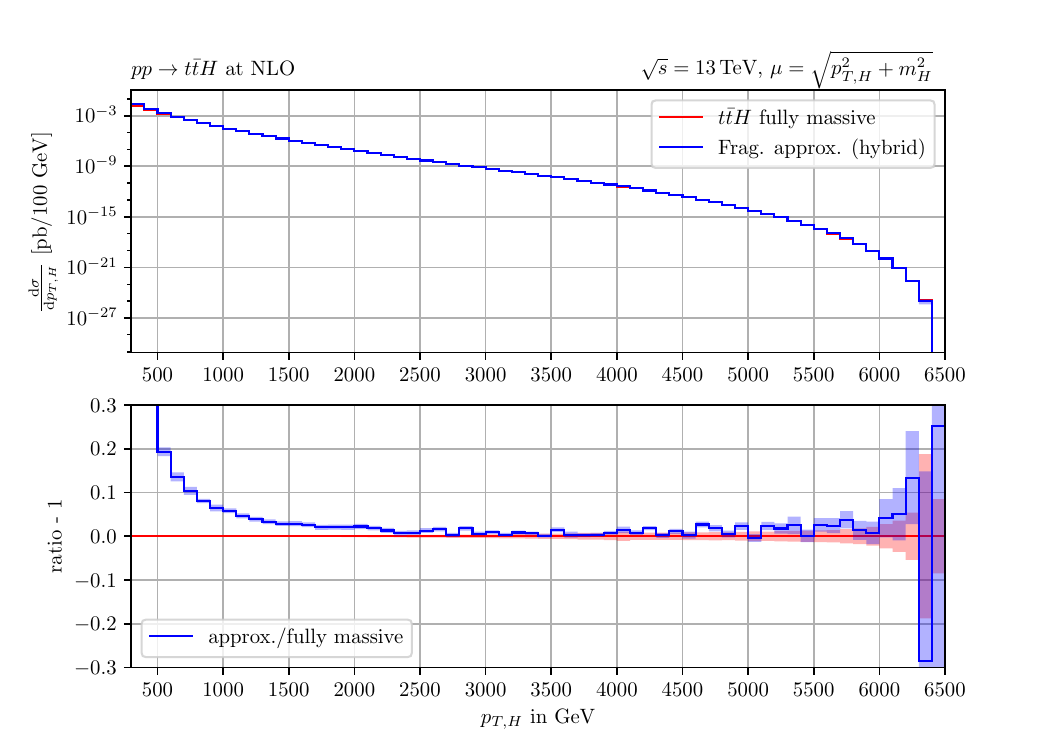}
    \caption{Same as Fig.~\ref{fig:LHC_Hybrid_NLO} but with all channels combined.}
    \label{fig:LHC_Hybrid_NLO_pp}
\end{figure}
Fig.~\ref{fig:LHC_Hybrid_NLO_pp} shows the combination of all channels (including the $qg$ channels). Above $p_T\approx2\,$TeV, the factorisation theorem can describe the fully-massive calculation reliably down to 1-2 \%. Below this, power corrections become more relevant and the two calculations start to differ more significantly.

Generally speaking, the factorisation theorem works reliably in the hybrid prescription up to NLO, even at LHC energies. This shows that the main power corrections in the factorisation theorem \eqref{eq:Masterformula} arise due to top-quark mass effects, while power corrections in the Higgs mass are less pronounced.

\section{Conclusions}
\label{sec:Conclusions}
We discussed and presented the first test of perturbative fragmentation functions for heavy non-QCD particles at NLO. We pointed out that, while the fully inclusive factorisation theorem is a straightforward generalisation of usual mass factorisation, the application to $t\tb H$ requiring a $t\tb$ pair in the final state becomes non-trivial starting at NNLO. Already at LO, we showed that the factorisation theorem in the general ZMTQ prescription cannot be used to reliably reproduce the massive $p_{T,H}$ spectrum in the $gg$ channel at LHC energies due to power corrections. We could however show that at a $100\,$TeV collider, the factorisation becomes viable. By including the leading top-mass dependence in the hybrid prescription, we found that the fragmentation formalism works very well in this prescription, even at LHC energies. This indicates that the power corrections of the ZMTQ prescription are mainly driven by top-quark mass effects.

\section*{Acknowledgements}
B.G. would like to thank Saimeng Zhou for useful discussions and for providing an analytic calculation of the fully-massive LO $gg\rightarrow t\tb H$ production cross section for some cross checks. C.B. acknowledges funding from the Italian Ministry of Universities and Research (MUR) through FARE grant R207777C4R. The work of M.C. was supported by the Deutsche Forschungsgemeinschaft (DFG) under grant 396021762 - TRR 257: Particle Physics Phenomenology after the Higgs Discovery. T.G. has been supported by STFC consolidated HEP theory grants ST/T000694/1 and ST/X000664/1. The work of B. G. has been supported by the DFG under grant 400140256 - GRK 2497: The physics of the heaviest particles at the Large Hadron Collider. Computations were performed with computing resources granted by the RWTH Aachen University under project p0020025.

\appendix
\section{Massless $t\tb H$ production at LO}
\label{app:masslessttH}
In this appendix, we present briefly LO results for the processes $q\qb\rightarrow t\tb H$ and $gg\rightarrow t\tb H$ with $m_t=m_H=0$. The calculations are performed in conventional dimensional regularisation (CDR) with $d=4-2\epsilon$ dimensions and are translated to the 't Hooft-Veltman scheme (tHV) used by \verb|STRIPPER|. Within the tHV scheme, resolved momenta are four-dimensional and only unresolved and loop momenta are treated in $d$ dimensions.

\subsection{Phase space parametrisation}
We start by writing the general 3-body phase space for the process
\begin{equation}
    a(p_a)+b(p_b)\rightarrow t(k)+\tb(\kb)+H(p)\;.
\end{equation}
After introducing a momentum $Q$ to split the phase space into one for the Higgs momentum $p$ and one for the $t\tb$ pair, there is
\begin{multline}
    \D\Phi=\left(\frac{\mu^4e^{2\gamma_E}}{(4\pi)^2}\right)^{\epsilon}\int\frac{\D^dp}{(2\pi)^d}(2\pi)\delta(p^2)\D^dQ\,\delta^{(d)}(p_a+p_b-p-Q)\\[0.1cm]
    \times\int\frac{\D^dk}{(2\pi)^d}(2\pi)\delta(k^2)\frac{\D^d\kb}{(2\pi)^d}(2\pi)\delta(\kb^2)(2\pi)^d\delta^{(d)}(Q-k-\kb)\;.
\end{multline}
The scale factor $\mu^{4\epsilon}$ ensures that the dimensionality of the phase space is correct. Because we include the factor $e^{2\gamma_E\epsilon}/(4\pi)^{2\epsilon}$, with Euler's gamma constant $\gamma_E$, minimal subtraction of the poles corresponds to the $\msb$ scheme.

Making use of the azimuthal symmetry of the process, we pick an explicit momentum parametrisation
\begin{align}
    p_a=&\;(\sqrt{s}/2,0,0,\sqrt{s}/2)^T,\\[0.2cm]
    p_b=&\;(\sqrt{s}/2,0,0,-\sqrt{s}/2)^T,\\[0.2cm]
    p=&\;(p_T\cosh{y},p_T,0,p_T\sinh{y})^T,
\end{align}
with $s$ the usual squared cms energy and $p_T$ and $y$ defining the transverse momentum and rapidity of the Higgs. The collinear limits can all be analysed using a fixed Higgs momentum, so we stay differential in the Higgs phase space. After performing the trivial azimuthal integrals and parametrising the Higgs phase space, the phase space is given by
\begin{equation}
    \frac{\D\Phi}{\D p_T^2\D y}=\frac{1}{64\pi^4\Gamma(1-\epsilon)}\left(\frac{p_T^2}{\pi e^{2\gamma_E}\mu^4}\right)^{-\epsilon}\int\D^dk\,\delta(k^2)\,\delta((Q-k)^2)\;,
\end{equation}
and $Q=(\sqrt{s}-p_T\cosh{y},-p_T,0,-p_T\sinh{y})^T$ is fixed. Notably, we parametrised $p_a$, $p_b$, $p$ and $Q$ in such a way that none of them depends on non-trivial $d-4$ dimensional components. In addition, scalar products of any momentum with one of these momenta cannot depend on such non-trivial components. Since $\kb$ is fixed by momentum conservation, every scalar product appearing in the matrix element contains one of the fixed momenta. Therefore, the integrals over the non-trivial $d-4$ dimensional components fully factorise and can be performed independently.

To parametrise the $t\tb$ phase space, we rotate to the frame in which $Q=(Q_0,0,0,\Qt)^T$ with $\Qt=-p_T\cosh{y}$, $Q_0=\sqrt{s}+\Qt$ and define $s_1\equiv Q^2=Q_0^2-\Qt^2$. This rotation allows a simple parametrisation of $k$ with respect to the axis defined by the Higgs momentum $p$ which makes parametrising the collinear limit of top and Higgs easy.

In the new frame, we define $k=(k_T\cosh{\eta},k_T\cos{\phi},k_T\sin{\phi},k_T\sinh{\eta})^T$. This yields
\begin{align}
    \notag
    \frac{\D\Phi}{\D p_T^2\D y}=&\;\frac{\sqrt{\pi}}{256\pi^4\,\Gamma(1-\epsilon)\Gamma(1/2-\epsilon)}\left(\frac{4p_T^2}{e^{2\gamma_E}\mu^4}\right)^{-\epsilon}\\[0.1cm]
    &\times\int\D k_T^2\D\eta\D\phi\,\left(k_T^2\sin^2{\phi}\right)^{-\epsilon}\,\delta\left(s_1-2\left[Q_0\cosh{\eta} + \Qt\sinh{\eta}\right]\right)\\[0.2cm]\notag
    =&\;\frac{1}{128\pi^4}\left(\frac{4p_T^2}{\mu^4}\right)^{-\epsilon}\int\frac{\D k_T^2\,\D\phi}{s_1\sqrt{1-\frac{4k_T^2}{s1}}}\left(k_T^2\sin^2{\phi}\right)^{-\epsilon} + \mathcal{O}(\epsilon^2)\\[0.1cm]
    =&\;\frac{1}{256\pi^4}\left(\frac{p_T^2\,s_1}{\mu^4}\right)^{-\epsilon}\int_0^1\D z\,\int_0^\pi\D\phi\,\left(1-z^2\right)^{-\epsilon}(\sin^2{\phi})^{-\epsilon}+ \mathcal{O}(\epsilon^2)\;,\label{eq:FinalPS}
\end{align}
where we used $z=\sqrt{1-\frac{4k_T^2}{s_1}}$ and expanded the $\Gamma$ functions in the second line up to $\mathcal{O}(\epsilon)$. Of the two possible solutions from the $\delta$-function, we picked the one with $\cosh{\eta}=(Q_0 + \Qt z)/(2k_T)$ and $\sinh{\eta}=(\Qt+Q_0 z)/(2k_T)$. This solution parametrises (in the limit $k_T\rightarrow0$) the collinear limit $k||p$. We can take the other solution\footnote{This would correspond to the limit $\kb||p$} into account by using the symmetrised matrix element
\begin{equation}
    \label{eq:symmetrisation}
    |\tilde{\mathcal{M}}|^2=|\mathcal{M}|^2(k,\kb)+|\mathcal{M}|^2(\kb,k)
\end{equation}
in the cross section calculation.

\subsection{$q\qb$ Channel}
In the $q\qb$ channel, the matrix element is given by two s-channel diagrams. Taking the symmetrisation of Eq.~\eqref{eq:symmetrisation} into account and plugging in the phase space parametrisation, we find
\begin{equation}
    |\tilde{\mathcal{M}}|^2=\frac{32\,y_t^2g^4}{s^2}\left[\left(\frac{1}{1-z}+\frac{1}{1+z}\right)[(s+s_1)+(s-s_1)\cos^2{\theta}]-2\frac{s_1}{1-z^2}\sin^2{\theta}-4\frac{s}{1-z^2}\epsilon\right]\;,
\end{equation}
which depends only on $z$ and not on $\phi$. Thus, the $z$ and $\phi$ integrals factorise and can be done independently of each other. Taking the flux factor and the colour and spin averages into account, we arrive at
\begin{equation}
    \frac{\D\sigma}{\D p_T^2\D y}= -\frac{y_t^2\alpha_S^2}{36\pi s^2}\left(\frac{p_T^2s_1}{\mu^4}\right)^{-\epsilon}(1+\cos^2{\theta})\frac{1}{\epsilon}+\frac{y_t\alpha_S^2}{18\pi s^2} + \mathcal{O}(\epsilon)\;,
\end{equation}
where the final term comes from the $\epsilon$ dependent part of the matrix element. The angle $\theta$ is the angle of the Higgs momentum relative to the beam axis. Since the fragmentation calculation performed in \verb|STRIPPER| uses the tHV scheme where $k$ is inherently 4-dimensional, we have to translate the CDR scheme of this calculation to the tHV scheme. This can be done by formulating the factorisation theorem in both schemes, i.e.\
\begin{align}\notag
    \D\sigma_{q\qb\rightarrow t\tb H}^{m_t\neq0,m_H=0}=&\left.\left[\D\sigma_{q\qb\rightarrow t\tb H}^{m_t=0,m_H=0}(p;\mu) + 2\int_0^1\D z\,\D\sigma_{q\qb\rightarrow t\tb}^{m_t=0}\left(k=\frac{p}{z};\mu\right)D^{m_t\neq 0,m_H=0}_{t\rightarrow H}(z;\mu)+\mathcal{O}(m_t^2)\right]\right|_{\text{CDR}}\\[0.2cm]
    =&\left.\left[\D\sigma_{q\qb\rightarrow t\tb H}^{m_t=0,m_H=0}(p;\mu) + 2\int_0^1\D z\,\D\sigma_{q\qb\rightarrow t\tb}^{m_t=0}\left(k=\frac{p}{z};\mu\right)D^{m_t\neq 0,m_H=0}_{t\rightarrow H}(z;\mu)+\mathcal{O}(m_t^2)\right]\right|_{\text{tHV}}.
    \label{eq:factorisation}
\end{align}
To infer the tHV formulation of the massless $t\tb H$ calculation, we perform the fragmentation calculation explicitly in the CDR (tHV) prescriptions, i.e.\ with $k$ in $d$ (four) dimensions. For both calculations, we use the same parametrisation of the $t\tb$ phase space as in the full $t\tb H$ calculation but make use of the fact that the rapidity of the top-quark is identical to the rapidity of the Higgs in the collinear limit. In addition, we use the unrenormalised version of the LO top-to-Higgs fragmentation function \cite{Braaten:2015ppa}. Comparing both calculations, we find
\begin{align}\notag
\label{eq:qqCDRFrag}
    \left.\frac{\D\sigma_{q\qb\rightarrow t(\rightarrow H)\tb}}{\D p_T^2\D y}\right|_{\text{CDR}}=&\left(\frac{p_T^2s^2}{(s-s_1)^2\mu^2}\right)^{-\epsilon}\left.\frac{\D\sigma_{q\qb\rightarrow t(\rightarrow H)\tb}}{\D p_T^2\D y}\right|_{\text{tHV}}-\frac{y_t^2\alpha_S^2}{36\pi s^2}\\[0.2cm]
    =&\;\frac{y_t^2\alpha_S^2}{72\pi s^2}(1+\cos^2{\theta})\left(\frac{1}{\epsilon}-\log{\frac{p_T^2m_t^2}{\mu^4}}+4\frac{ss_1}{(s-s_1)^2}\right) -\frac{y_t^2\alpha_S^2}{36\pi s^2}+\mathcal{O}(\epsilon)\;,
\end{align}
where we defined the short hand
\begin{equation}
    \frac{\D\sigma_{q\qb\rightarrow t(\rightarrow H)\tb}}{\D p_T^2\D y}=\int_0^1\D z\,\frac{\D\sigma_{q\qb\rightarrow t\tb}^{m_t=0}}{\D p_T^2 \D y}\left(k=\frac{p}{z};\mu\right)D^{m_t\neq 0,m_H=0}_{t\rightarrow H}(z;\mu)\;.
\end{equation}
The final term in Eq.~\eqref{eq:qqCDRFrag} appears because the matrix element in the CDR calculation depends explicitly on $\epsilon$ which is not the case for the tHV calculation\footnote{This is an artifact of the implementation of the tHV scheme within STRIPPER where no $\mathcal{O}(\epsilon)$ terms of tree level matrix elements are required \cite{Czakon:2019tmo}. Because the goal is to match the massless calculation to a fragmentation calculation done via STRIPPER, we have to match this version of the tHV scheme as well.}. Using the factorisation theorem, we find
\begin{align}\notag
    \left.\frac{\D\sigma_{q\qb\rightarrow t\tb H}^{m_t=0,m_H=0}}{\D p_T^2\D y}(p;\mu)\right|_{\text{tHV}}=&\left.\frac{\D\sigma_{q\qb\rightarrow t\tb H}^{m_t=0,m_H=0}}{\D p_T^2\D y}(p;\mu)\right|_{\text{CDR}}+\left.\frac{\D\sigma_{q\qb\rightarrow t(\rightarrow H)\tb}}{\D p_T^2\D y}\right|_{\text{CDR}}-\left.\frac{\D\sigma_{q\qb\rightarrow t(\rightarrow H)\tb}}{\D p_T^2\D y}\right|_{\text{tHV}}\\[0.2cm]
    =&-\frac{y_t^2\alpha_S^2}{36\pi s^2}\left(\frac{(s-s_1)^2s_1}{s^2\mu^2}\right)^{-\epsilon}(1+\cos^2\theta)\frac{1}{\epsilon}+\mathcal{O}(\epsilon)\;.
\end{align}
Note that the contributions from the $\epsilon$ dependence of the matrix element cancel. To simplify the notation, we rewrite the combined logarithms as a power in $\epsilon$ making the result accurate through $\mathcal{O}(\epsilon^0)$. We cross checked explicitly that, when combined with the tHV fragmentation calculation, the correct logarithmic dependence of the massive cross section is obtained which can be inferred from Appendix B of Ref.~\cite{Braaten:2015ppa}.

\subsection{$gg$ Channel}
In the $gg$ channel, $t$ and $u$ channel diagrams appear. These lead to divergences if a massless top or anti-top-quark becomes collinear to an initial-state gluon. Within the given phase space parametrisation, this leads to diverging propagators depending on both $z$ and $\phi$ which means that the integrals do not factorise. This makes a fully analytic calculation very difficult. The goal is therefore to bring the integrand into a form that allows for numerical integration at fixed Higgs momentum. For differential distributions related to the Higgs, this procedure leads to much more stable results than the scalar subtraction scheme (see Appendix~\ref{app:ScalarSubtraction}) because missed-binning effects are avoided by construction.

The phase space measure in Eq.~\eqref{eq:FinalPS} is modified slightly by requiring $y>0$ and by noticing that the matrix element in the $gg$ channel is inherently symmetric under the exchange $k\leftrightarrow \kb$ which means that no symmetrisation is needed. Together, this leads to an additional factor 4 in the phase space measure compared to Eq.~\eqref{eq:FinalPS}.

We define the following kinematic invariants for the problem:
\begin{align}
    s_{12}=&\;(p+k)^2,\,\qquad s_{23}=(p+\kb)^2,\;\;\qquad t_1=(p_b-\kb)^2,\\[0.2cm]
    t_2=&\;(p_a-k)^2,\qquad u_1=(p_a-\kb)^2,\qquad u_2=(p_b-k)^2\;.
\end{align}
It is clear that the exchange $k\leftrightarrow\kb$ corresponds to the changes $s_{12}\leftrightarrow s_{23}$, $t_1\leftrightarrow u_2$ and $t_2\leftrightarrow u_1$. Within the phase space parametrisation, this symmetry can be expressed by the joint exchanges $z\leftrightarrow-z$ and $\phi\leftrightarrow\phi+\pi$.

The cross section is logarithmically divergent in the limit of any of the kinematic invariants being zero. Since we are interested in the high $p_T$ behaviour of the Higgs, none of these limits can overlap. We are interested in the poles of the propagators, i.e.\ the roots of the kinematic invariants in the given phase space parametrisation. By construction of the phase space, the invariants $s_{12}$ and $s_{23}$ are linear in $z$ and have single roots at $z=1$ and $z=-1$, independent of $\phi$. The other kinematic invariants are more complicated. They are given by
\begin{align}
    t_{1}&=-\frac{\sqrt{s}}{2}\left(Q_0 - \Qt z + (\Qt - Q_0 z)\cos{\theta} + \sqrt{s_1(1-z^2)}\sin{\theta}\cos{\phi}\right),\\[0.2cm]
    t_{2}&=-\frac{\sqrt{s}}{2}\left(Q_0 + \Qt z - (\Qt + Q_0 z)\cos{\theta} + \sqrt{s_1(1-z^2)}\sin{\theta}\cos{\phi}\right),\\[0.2cm]
    u_{1}&=-\frac{\sqrt{s}}{2}\left(Q_0 - \Qt z - (\Qt - Q_0 z)\cos{\theta} - \sqrt{s_1(1-z^2)}\sin{\theta}\cos{\phi}\right),\\[0.2cm]
    u_{2}&=-\frac{\sqrt{s}}{2}\left(Q_0 + \Qt z + (\Qt + Q_0 z)\cos{\theta} - \sqrt{s_1(1-z^2)}\sin{\theta}\cos{\phi}\right),
\end{align}
and depend non-rationally on $z$. Their roots are given by
\begin{align}
    z_{t_1}&=\frac{\Qt + Q_0\cos{\theta}}{Q_0+\Qt\cos{\theta}}\;,\;\;\,\quad\phi=\pi\;,\\[0.2cm]
    z_{t_2}&=-\frac{\Qt-Q_0\cos{\theta}}{Q_0-\Qt\cos{\theta}}\;,\quad\phi=\pi\;,\\[0.2cm]
    z_{u_1}&=\frac{\Qt-Q_0\cos{\theta}}{Q_0-\Qt\cos{\theta}}\;,\;\;\,\quad\phi=0\;,\\[0.2cm]
    z_{t_1}&=-\frac{\Qt + Q_0\cos{\theta}}{Q_0+\Qt\cos{\theta}}\;,\quad\phi=0\;.
\end{align}
It should be pointed out that not all of the roots of the invariants lie within the integration range $z\in[0,1]$. This is most apparent for the $z=-1$ root of $s_{23}$ but can be the case for the other roots as well, depending on the Higgs momentum. We decide to treat these roots on the same footing as the roots that lie within the integration range. This is possible because the residues of the matrix element stay regular for $z\in[-1,0]$. Including the terms related to roots outside the integration range corresponds therefore to adding and subtracting a finite term in the squared matrix element. The advantage is that we do not need to take care of the different regions of the Higgs phase space where different roots of the kinematic invariants lie within the integration range. Also, it makes the symmetry under $k\leftrightarrow\kb$ manifest in all terms making it easier to integrate the subtraction terms.

Treating the matrix element as a function of $z$ and $\phi$, we can define the ad hoc subtractions
\begin{align}
    \notag
    \M(z,\phi)=\biggl\{\M(z,\phi)&-\left[\M(z,\phi)s_{12}\right]_{z=1}\frac{1}{s_{12}}-\left[\M(z,\phi)s_{23}\right]_{z=-1}\frac{1}{s_{23}}\\[0.2cm]\notag
    &-\left[\M(z,\phi)t_1\right]_{z=z_{t_1},\phi=\pi}\frac{1}{t_1}-\left[\M(z,\phi)t_2\right]_{z=z_{t_2},\phi=\pi}\frac{1}{t_2}\\[0.2cm]
    &-\left[\M(z,\phi)u_1\right]_{z=z_{u_1},\phi=0}\frac{1}{u_1}\left.-\left[\M(z,\phi)u_2\right]_{z=z_{u_2},\phi=0}\frac{1}{u_2}\right\}\\[0.2cm]\notag
    &+\left[\M(z,\phi)s_{12}\right]_{z=1}\frac{1}{s_{12}}+\left[\M(z,\phi)s_{23}\right]_{z=-1}\frac{1}{s_{23}}\\[0.2cm]\notag
    &+\left[\M(z,\phi)t_1\right]_{z=z_{t_1},\phi=\pi}\frac{1}{t_1}+\left[\M(z,\phi)t_2\right]_{z=z_{t_2},\phi=\pi}\frac{1}{t_2}\\[0.2cm]\notag
    &+\left[\M(z,\phi)u_1\right]_{z=z_{u_1},\phi=0}\frac{1}{u_1}+\left[\M(z,\phi)u_2\right]_{z=z_{u_2},\phi=0}\frac{1}{u_2}\;.
\end{align}
By construction, the combination in the braces is finite and can be integrated over $z$ and $\phi$ numerically in four dimensions without missed-binning effects. The terms outside the braces will be referred to as the integrated subtraction terms. Due to the symmetry of the matrix element under $k\leftrightarrow\kb$, the matrix element evaluated at the roots of the kinematic invariants can be related yielding
\begin{align}
    \notag
    \M(z,\phi)=&\;\biggl\{\M(z,\phi)-\left[\M(z,\phi)s_{12}\right]_{z=1}\left(\frac{1}{s_{12}}+\frac{1}{s_{23}}\right)\\[0.2cm]\notag
    &-\left[\M(z,\phi)t_2\right]_{z=z_{t_2},\phi=\pi}\left(\frac{1}{t_2}+\frac{1}{u_1}\right)-\left[\M(z,\phi)t_1\right]_{z=z_{t_1},\phi=\pi}\left(\frac{1}{t_1}+\frac{1}{u_2}\right)\biggr\}\\[0.2cm]
    &+\left[\M(z,\phi)s_{12}\right]_{z=1}\left(\frac{1}{s_{12}}+\frac{1}{s_{23}}\right)\\[0.2cm]\notag
    &+\left[\M(z,\phi)t_2\right]_{z=z_{t_2},\phi=\pi}\left(\frac{1}{t_2}+\frac{1}{u_1}\right)+\left[\M(z,\phi)t_1\right]_{z=z_{t_1},\phi=\pi}\left(\frac{1}{t_1}+\frac{1}{u_2}\right).
    \label{eq:subtractions}
\end{align}
The integrated subtraction term for the final state collinear divergence between the Higgs and the top or anti-top-quark can be evaluated straightforwardly in the given phase space parametrisation:
\begin{align}
    \notag
    \int\frac{\D\Phi}{\D p_T^2\D y\D\phi}\left(\frac{1}{s_{12}} + \frac{1}{s_{23}}\right)&=\frac{1}{64\pi^4}\left(\frac{p_T^2s_1\sin^2{\phi}}{\mu^4}\right)^{-\epsilon}\frac{2}{-\Qt(Q_0+\Qt)}\int_0^1\frac{\D z}{(1-z^2)^{1+\epsilon}}\\[0.2cm]
    &=\frac{1}{64\pi^4}\left(\frac{p_T^2s_1\sin^2{\phi}}{\mu^4}\right)^{-\epsilon}\frac{1}{\Qt(Q_0+\Qt)}\frac{1}{\epsilon}\;.
\end{align}
Notably, the $z$ and $\phi$ integrals factorise. Also, since the roots of the denominator are independent of $\phi$, the residue still has to be integrated over $\phi$ which can however be done numerically.

For the integrated subtraction terms related to the $t$ and $u$ channels, a straightforward evaluation is not possible because the $z$ and $\phi$ integrals interfere non-trivially with each other and additional regularisations would be necessary for this parametrisation of the $t\tb$ phase space. The introduction of the subtraction terms however isolates the different collinear limits. Therefore, it is possible to pick a new phase space parametrisation, once for the $t_1$, $u_2$ integrated subtraction term and once for the $t_2$, $u_1$ integrated subtraction term. We do this by fully reparametrising the $t\tb$ phase space which can be done due to the $k\leftrightarrow\kb$ symmetry still inherent in the integrated subtraction terms and the fact that the residues of these terms are fixed. Keeping the Higgs phase space as it is, we boost to the cms-frame of the $t\tb$ pair and rotate the frame such that $p_a$ ($p_b$) points in the $x_3$-direction of the new frame for the integrated subtraction term involving $t_2$, $u_1$ ($t_1$, $u_2$). Relative to this axis, we can parametrise the $t\tb$ phase space again by $k_T$, rapidity $\eta$ and azimuthal angle $\phi$ in the same way as for the parametrisation relative to the Higgs momentum axis. In this frame, the integrals over $z=\sqrt{1-\frac{4k_T^2}{s_1}}$ and $\phi$ factorise and can be done independently yielding
\begin{align}
    \int\frac{\D\Phi}{\D p_T^2\D y}\left(\frac{1}{t_2}+\frac{1}{u_1}\right)=&\;\frac{1}{32\pi^3}\frac{1}{\sqrt{s}(Q_0-\Qt\cos{\theta})}\left(\frac{p_T^2s_1}{\mu^4}\right)^{-\epsilon}\frac{1}{\epsilon}\;,\\[0.2cm]
    \int\frac{\D\Phi}{\D p_T^2\D y}\left(\frac{1}{t_1}+\frac{1}{u_2}\right)=&\;\frac{1}{32\pi^3}\frac{1}{\sqrt{s}(Q_0+\Qt\cos{\theta})}\left(\frac{p_T^2s_1}{\mu^4}\right)^{-\epsilon}\frac{1}{\epsilon}\;,
\end{align}
for the two remaining integrated subtraction terms. The first term takes the collinear limits of a top or anti-top relative to the first initial-state gluon of the process $gg\rightarrow t\tb H$ into account, the second term the limits relative to the second initial-state gluon.

To match this calculation to the tHV scheme, we can use the same procedure as for the $q\qb$ channel. We write the factorisation theorem as
\begin{align}\notag
    \label{eq:factorisationGG}
    \D\sigma_{gg\rightarrow t\tb H}^{m_t\neq0,m_H=0}=&\left[\D\sigma_{q\qb\rightarrow t\tb H}^{m_t=0,m_H=0}(p;\mu) 
    +\D\sigma_{gg\rightarrow t(\rightarrow H)\tb}
    +\D\sigma_{gg\rightarrow t\tb(\rightarrow H)}\right.\\[0.1cm]
    &+f_{g\rightarrow t}\otimes\D\sigma_{tg\rightarrow tH}
    +f_{g\rightarrow \tb}\otimes\D\sigma_{\tb g\rightarrow \tb H}\\[0.1cm]\notag
    &\left.\left.+f_{g\rightarrow t}\otimes\D\sigma_{gt\rightarrow tH}
    +f_{g\rightarrow \tb}\otimes\D\sigma_{g\tb\rightarrow \tb H}\right]\right|_{\text{CDR, tHV}}\;,
\end{align}
where we wrote out all of the terms coming from the limiting behaviour of each collinear limit explicitly.

The scheme change can thus be written as
\begin{align}
    \notag
    \left.\frac{\D\sigma_{gg\rightarrow t\tb H}^{m_t=0,m_H=0}}{\D p_T^2\D y}(p;\mu)\right|_{\text{tHV}}=&\left.\frac{\D\sigma_{gg\rightarrow t\tb H}^{m_t=0,m_H=0}}{\D p_T^2\D y}(p;\mu)\right|_{\text{CDR}}\\[0.2cm]\notag
    &+\Delta\frac{\D\sigma_{gg\rightarrow t(\rightarrow H)\tb}}{\D p_T^2\D y}(p;\mu)
    +\Delta\frac{\D\sigma_{gg\rightarrow t\tb(\rightarrow H)}}{\D p_T^2\D y}(p;\mu)\\[0.2cm]
    \label{eq:ggRelationCDRtHV}
    &+\Delta\left(f_{g\rightarrow t}\otimes\frac{\D\sigma_{tg\rightarrow tH}}{\D p_T^2\D y}(p;\mu)\right)
    +\Delta\left(f_{g\rightarrow \tb}\otimes\frac{\D\sigma_{\tb g\rightarrow \tb H}}{\D p_T^2\D y}(p;\mu)\right)\\[0.2cm]\notag
    &+\Delta\left(f_{g\rightarrow t}\otimes\frac{\D\sigma_{gt\rightarrow tH}}{\D p_T^2\D y}(p;\mu)\right)
    +\Delta\left(f_{g\rightarrow \tb}\otimes\frac{\D\sigma_{g\tb\rightarrow \tb H}}{\D p_T^2\D y}(p;\mu)\right),
\end{align}
where $\Delta$ denotes the difference between the CDR and the tHV version of the corresponding quantities. For the terms related to the final state, we expect the same behaviour as in the $q\qb$ channel, i.e.
\begin{multline}
    \Delta\frac{\D\sigma_{gg\rightarrow t(\rightarrow H)\tb}}{\D p_T^2\D y}(p;\mu)
    +\Delta\frac{\D\sigma_{gg\rightarrow t\tb(\rightarrow H)}}{\D p_T^2\D y}(p;\mu)
    =-\frac{1}{512s}\frac{1}{64\pi^4}\frac{1}{\Qt(Q_0+\Qt)}\\[0.1cm]\times\int_0^{\pi}\D\phi\,\left[\frac{\left[\M(z,\phi)s_{12}\right]_{z=1}}{(1-\epsilon)^2}\left(\frac{p_T^2s^2\sin^2{\phi}}{(s-s_1)^2\mu^2}\right)^{-\epsilon}-\left[\M(z,\phi)s_{12}\right]_{z=1,\epsilon=0}\right]\frac{1}{\epsilon}+\mathcal{O}(\epsilon)\;.
\end{multline}
The factor $1/(1-\epsilon)^2$ arises because the gluon spin average is done with $d-2$ spin states in CDR. The residue of the matrix element is evaluated in CDR and contains an $\mathcal{O}(\epsilon)$ term\footnote{As for the $q\qb$ channel, there are no $\mathcal{O}(\epsilon)$ terms in the matrix element of the tHV scheme due to the implementation within STRIPPER.}.

The terms in Eq.~\eqref{eq:ggRelationCDRtHV} related to the integrated subtraction term with $t_2$ and $u_1$ are
\begin{align}\notag
    \Delta&\left(f_{g\rightarrow t}\otimes\frac{\D\sigma_{tg\rightarrow tH}}{\D p_T^2\D y}(p;\mu)\right)
    +\Delta\left(f_{g\rightarrow \tb}\otimes\frac{\D\sigma_{\tb g\rightarrow \tb H}}{\D p_T^2\D y}(p;\mu)\right)
    \\[0.2cm]\notag&=-\frac{1}{512s}\frac{\pi}{64\pi^4}\left[\frac{\left[\M(z,\phi)t_2\right]_{z=z_{t_2},\phi=\pi}}{\sqrt{s}(Q_0+\Qt\cos{\theta})(1-\epsilon)}\left(\frac{p_T^2}{\mu^2}\right)^{-\epsilon}-\frac{\left[\M(z,\phi)t_2\right]_{z=z_{t_2},\phi=\pi,\epsilon=0}}{\sqrt{s}(Q_0+\Qt\cos{\theta})}\right]\frac{2}{\epsilon}+\mathcal{O}(\epsilon)
    \\[0.2cm]&=-\frac{1}{512s}\frac{\pi}{64\pi^4}\frac{\left[\M(z,\phi)t_2\right]_{z=z_{t_2},\phi=\pi,\epsilon=0}}{\sqrt{s}(Q_0+\Qt\cos{\theta})}\left[\left(\frac{p_T^2}{\mu^2}\right)^{-\epsilon}-1\right]\frac{2}{\epsilon}+\mathcal{O}(\epsilon)\;.
\end{align}
From the second to the third line, we used the fact that the residue of the matrix element in CDR is proportional to $(1-\epsilon)$ which cancels the factor coming from the gluon spin average in CDR. Similarly, we find for the terms related to the integrated subtraction term with $t_1$ and $u_2$
\begin{align}\notag
    \Delta&\left(f_{g\rightarrow t}\otimes\frac{\D\sigma_{gt\rightarrow tH}}{\D p_T^2\D y}(p;\mu)\right)
    +\Delta\left(f_{g\rightarrow \tb}\otimes\frac{\D\sigma_{g\tb\rightarrow \tb H}}{\D p_T^2\D y}(p;\mu)\right)
    \\[0.2cm]\notag&=-\frac{1}{512s}\frac{\pi}{64\pi^4}\left[\frac{\left[\M(z,\phi)t_1\right]_{z=z_{t_1},\phi=\pi}}{\sqrt{s}(Q_0-\Qt\cos{\theta})(1-\epsilon)}\left(\frac{p_T^2}{\mu^2}\right)^{-\epsilon}-\frac{\left[\M(z,\phi)t_1\right]_{z=z_{t_1},\phi=\pi,\epsilon=0}}{\sqrt{s}(Q_0-\Qt\cos{\theta})}\right]\frac{2}{\epsilon}+\mathcal{O}(\epsilon)
    \\[0.2cm]&=-\frac{1}{512s}\frac{\pi}{64\pi^4}\frac{\left[\M(z,\phi)t_1\right]_{z=z_{t_1},\phi=\pi,\epsilon=0}}{\sqrt{s}(Q_0-\Qt\cos{\theta})}\left[\left(\frac{p_T^2}{\mu^2}\right)^{-\epsilon}-1\right]\frac{2}{\epsilon}+\mathcal{O}(\epsilon)\;.
\end{align}
Inserting these terms in Eq.~\eqref{eq:ggRelationCDRtHV} and combining them through $\mathcal{O}(\epsilon^0)$, we find
\begin{align}
    \notag
    \label{eq:ggMasslessResult}
    \left.\frac{\D\sigma_{gg\rightarrow t\tb H}^{m_t=0,m_H=0}}{\D p_T^2\D y}\right|_{\text{tHV}}=&\;\frac{1}{512s}\frac{1}{64\pi^4}\biggl\{\int_0^1\D z\int_0^{\pi}\D\phi\,\left[\M(z,\phi)-\left[\M(z,\phi)s_{12}\right]_{z=1}\left(\frac{1}{s_{12}} + \frac{1}{s_{23}}\right)\right.\\[0.2cm]\notag
    &\qquad\qquad\qquad-\left[\M(z,\phi)t_2\right]_{z=z_{t_2},\phi=\pi}\left(\frac{1}{t_2}+\frac{1}{u_1}\right)\\[0.2cm]\notag
    &\qquad\qquad\qquad\left.-\left[\M(z,\phi)t_1\right]_{z=z_{t_1},\phi=\pi}\left(\frac{1}{t_1}+\frac{1}{u_2}\right)\right]_{\epsilon=0}\\[0.2cm]
    &\qquad +\int_0^{\pi}\D\phi\,\frac{\left[\M(z,\phi)s_{12}\right]_{z=1,\epsilon=0}}{\Qt(Q_0+\Qt)}\left(\frac{(s-s_1)^2s_1}{s^2\mu^2}\right)^{-\epsilon}\frac{1}{\epsilon}\\[0.2cm]\notag
    &\qquad +\frac{2\pi}{(1-\epsilon)^2}\frac{\left[\M(z,\phi)t_2\right]_{z=z_{t_2},\phi=\pi}}{\sqrt{s}(Q_0-\Qt\cos{\theta})}\left(\frac{s_1}{\mu^2}\right)^{-\epsilon}\frac{1}{\epsilon}\\[0.2cm]\notag
    &\qquad +\frac{2\pi}{(1-\epsilon)^2}\frac{\left[\M(z,\phi)t_2\right]_{z=z_{t_2},\phi=\pi}}{\sqrt{s}(Q_0+\Qt\cos{\theta})}\left(\frac{s_1}{\mu^2}\right)^{-\epsilon}\frac{1}{\epsilon}\biggr\}+\mathcal{O}(\epsilon)\;.
\end{align}
The residue related to $s_{12}$ is evaluated at $\epsilon=0$ because the $\mathcal{O}(\epsilon)$ terms cancel against the same terms arising in the scheme change of the fragmentation calculation. The same holds true for the gluon spin average in this term. Such cancellations do not appear for the other integrated subtraction terms because here, the $\epsilon$ dependence of the matrix element cancels the dependence from the gluon spin average in the scheme change terms. Therefore, the residues in the final two lines have to be evaluated at $\epsilon\neq0$ and the gluon spin average of CDR has to be taken into account. By dropping and introducing some $\mathcal{O}(\epsilon)$ terms, the combination of the different terms in Eq.~\eqref{eq:ggRelationCDRtHV} yields the more concise form of Eq.~\eqref{eq:ggMasslessResult} which is accurate through $\mathcal{O}(\epsilon^0)$.

As a check, we reconstructed the logarithmic dependence of the $gg\rightarrow t\tb H$ calculation with $m_t\neq0$, $m_H=0$. To do this, we applied the same subtraction procedure to the massive calculation as for the massless calculation and expanded the integrands in $m_t$. The integration boundaries of the phase space parametrisation can be expanded in the regular part but have to be kept in the integrated subtraction terms. These can be evaluated yielding the logarithmic dependence in a form that can be compared directly to Eq.~\eqref{eq:factorisationGG}.

The remaining integrals in Eq.~\eqref{eq:ggMasslessResult} are all finite and can be performed numerically. Notably, the subtractions are fully local in the Higgs phase space, so no missed-binning effects can appear for the observables considered in this paper. To obtain the matrix elements, we used \verb|FORM| \cite{Vermaseren:2000nd,Kuipers:2012rf}. For the phase space integration and the convolution with the PDFs, a very basic Monte Carlo program was used. Compared to the evaluation of the massless cross sections with the scalar subtraction scheme described in Appendix~\ref{app:ScalarSubtraction}, about a factor $10^2$ less points were required at about a factor $7.5$ less computation time per point for a similar precision.

\section{A scalar subtraction scheme for STRIPPER}
\label{app:ScalarSubtraction}
We extend the implementation of the \verb|STRIPPER| library to accommodate unresolved massless scalar particles with a massless quark as the reference particle at NLO. For this, we define new sectors specific to this setup. Their selector function is the same as the one of a $q\qb$ sector (see Ref.~\cite{Czakon:2014oma}). Compared to pure QCD sectors, no divergences are included for a soft scalar particle because we will only be interested in the production of a scalar at finite transverse momentum. Therefore, we only include collinear subtractions. \verb|STRIPPER| is well suited for this because soft, collinear and soft-collinear subtractions are treated independently and the integration over the unresolved phase space of the integrated subtraction terms is done numerically. Hence, the soft and soft-collinear subtractions can be deactivated easily for scalar sectors and the quark-to-scalar splitting function can be included for the collinear subtraction without the need for any analytical integration.

The splitting function can be obtained from the explicit factorisation of a matrix element. A general matrix element for producing a quark (in this context a top-quark) which radiates off a collinear scalar particle (in this context a Higgs) can be written generically as
\begin{equation}
    |\mathcal{M}_{tH}|^2\approx\frac{2y_t^2}{(u+r)^2}\bra{\mathcal{M}^{s}_t}P^{ss'}_{t\rightarrow H}(z)\ket{\mathcal{M}^{s'}_{t}}\;,
\end{equation}
where $r$ ($u$) is the (un-)resolved momentum and $z=u_0/(u_0+r_0)$ is the momentum fraction carried by the unresolved particle. $\ket{\mathcal{M}_t^s}$ denotes the matrix element where the Higgs boson has been removed and the top-quark has spin $s$. Explicit evaluation gives $P^{ss'}_{t\rightarrow H}(z)=\delta^{ss'}z/2$. Averaged over the spin of the top-quark, we find $P_{t\rightarrow H}(z)=z/2$ which agrees with the result given in Ref.~\cite{Brancaccio:2021gcz} after accounting for varying conventions.

The subtraction scheme was verified against the semi-analytic calculations of $q\qb\rightarrow t\tb H$ and $gg\rightarrow t\tb H$ presented in Appendix~\ref{app:masslessttH} by checking that both the poles and the finite part of the cross section match within the errors which were on the order of $1.5\%$ for most bins of the subtraction scheme calculation and below $1\%$ for the semi-analytic calculation. Also, we performed the massless calculation using this scalar subtraction scheme to repeat the LO analysis in Sec.~\ref{sec:ResultsZMTQ}, yielding the same results.

\bibliographystyle{JHEP}
\bibliography{bibliography}

@article{ATLAS:2024gth,
    author = "Aad, Georges and others",
    collaboration = "ATLAS",
    title = "{Measurement of the associated production of a top-antitop-quark pair and a Higgs boson decaying into a $b\bar{b}$ pair in pp collisions at $\sqrt{s}=13$~TeV using the ATLAS detector at the LHC}",
    eprint = "2407.10904",
    archivePrefix = "arXiv",
    primaryClass = "hep-ex",
    reportNumber = "CERN-EP-2024-194",
    doi = "10.1140/epjc/s10052-025-13740-x",
    journal = "Eur. Phys. J. C",
    volume = "85",
    number = "2",
    pages = "210",
    year = "2025"
}

@article{CMS:2024fdo,
    author = "Hayrapetyan, Aram and others",
    collaboration = "CMS",
    title = "{Measurement of the $ \textrm{t}\overline{\textrm{t}}\textrm{H} $ and tH production rates in the H {\textrightarrow}$ \textrm{b}\overline{\textrm{b}} $ decay channel using proton-proton collision data at $ \sqrt{s} $ = 13 TeV}",
    eprint = "2407.10896",
    archivePrefix = "arXiv",
    primaryClass = "hep-ex",
    reportNumber = "CMS-HIG-19-011, CERN-EP-2024-179",
    doi = "10.1007/JHEP02(2025)097",
    journal = "JHEP",
    volume = "02",
    pages = "097",
    year = "2025"
}

@article{Balsach:2025jcw,
    author = "Balsach, Roger and others",
    title = "{State-of-the-art cross sections for $t\bar t H$: NNLO predictions matched with NNLL resummation and EW corrections}",
    eprint = "2503.15043",
    archivePrefix = "arXiv",
    primaryClass = "hep-ph",
    reportNumber = "LHCHWG-2025-001, IPPP/25/01, TUM-HEP-1549/25, UWThPh 2024-25, TTK-25-01, P3H-25-001, TIF-UNIMI-2025-3, MS-TP-25-02, PSI-PR-25-03, ZU-TH 08/25",
    doi = "10.21468/SciPostPhysCommRep.10",
    month = "3",
    year = "2025",
    journal={SciPost Phys. Comm. Rep.},
	pages={10},
}

@article{Beenakker:2001rj,
    author = "Beenakker, W. and Dittmaier, S. and Kr{\"a}mer, M. and Plumper, B. and Spira, M. and Zerwas, P. M.",
    title = "{Higgs radiation off top quarks at the Tevatron and the LHC}",
    eprint = "hep-ph/0107081",
    archivePrefix = "arXiv",
    reportNumber = "DESY-01-077, EDINBURGH-2001-08, PSI-PR-01-10",
    doi = "10.1103/PhysRevLett.87.201805",
    journal = "Phys. Rev. Lett.",
    volume = "87",
    pages = "201805",
    year = "2001"
}

@article{Reina:2001sf,
    author = "Reina, L. and Dawson, S.",
    title = "{Next-to-leading order results for t anti-t h production at the Tevatron}",
    eprint = "hep-ph/0107101",
    archivePrefix = "arXiv",
    reportNumber = "FSU-HEP-2001-0601, BNL-HET-01-20",
    doi = "10.1103/PhysRevLett.87.201804",
    journal = "Phys. Rev. Lett.",
    volume = "87",
    pages = "201804",
    year = "2001"
}

@article{Reina:2001bc,
    author = "Reina, L. and Dawson, S. and Wackeroth, D.",
    title = "{QCD corrections to associated t anti-t h production at the Tevatron}",
    eprint = "hep-ph/0109066",
    archivePrefix = "arXiv",
    reportNumber = "FSU-HEP-2001-0602, BNL-HET-01-19, UR-1639",
    doi = "10.1103/PhysRevD.65.053017",
    journal = "Phys. Rev. D",
    volume = "65",
    pages = "053017",
    year = "2002"
}

@article{Beenakker:2002nc,
    author = "Beenakker, W. and Dittmaier, S. and Kr{\"a}mer, M. and Plumper, B. and Spira, M. and Zerwas, P. M.",
    title = "{NLO QCD corrections to t anti-t H production in hadron collisions}",
    eprint = "hep-ph/0211352",
    archivePrefix = "arXiv",
    reportNumber = "DESY-02-177, EDINBURGH-2002-18, MPI-PHT-2002-70, PSI-PR-02-22",
    doi = "10.1016/S0550-3213(03)00044-0",
    journal = "Nucl. Phys. B",
    volume = "653",
    pages = "151--203",
    year = "2003"
}

@article{Dawson:2003zu,
    author = "Dawson, S. and Jackson, C. and Orr, L. H. and Reina, L. and Wackeroth, D.",
    title = "{Associated Higgs production with top quarks at the large hadron collider: NLO QCD corrections}",
    eprint = "hep-ph/0305087",
    archivePrefix = "arXiv",
    reportNumber = "BNL-HET-03-9, FSU-HEP-2003-0503, UB-HET-03-02",
    doi = "10.1103/PhysRevD.68.034022",
    journal = "Phys. Rev. D",
    volume = "68",
    pages = "034022",
    year = "2003"
}

@article{Brancaccio:2021gcz,
    author = {Brancaccio, Colomba and Czakon, Micha{\l} and Generet, Terry and Kr{\"a}mer, Michael},
    title = "{Higgs-boson production in top-quark fragmentation}",
    eprint = "2106.06516",
    archivePrefix = "arXiv",
    primaryClass = "hep-ph",
    reportNumber = "P3H-21-046, TTK-21-21",
    doi = "10.1007/JHEP08(2021)145",
    journal = "JHEP",
    volume = "08",
    pages = "145",
    year = "2021"
}

@article{Frixione:2014qaa,
    author = "Frixione, S. and Hirschi, V. and Pagani, D. and Shao, H. S. and Zaro, M.",
    title = "{Weak corrections to Higgs hadroproduction in association with a top-quark pair}",
    eprint = "1407.0823",
    archivePrefix = "arXiv",
    primaryClass = "hep-ph",
    reportNumber = "CERN-PH-TH-2014-123, CP3-14-49",
    doi = "10.1007/JHEP09(2014)065",
    journal = "JHEP",
    volume = "09",
    pages = "065",
    year = "2014"
}

@article{Zhang:2014gcy,
    author = "Zhang, Yu and Ma, Wen-Gan and Zhang, Ren-You and Chen, Chong and Guo, Lei",
    title = "{QCD NLO and EW NLO corrections to $t\bar{t}H$ production with top quark decays at hadron collider}",
    eprint = "1407.1110",
    archivePrefix = "arXiv",
    primaryClass = "hep-ph",
    doi = "10.1016/j.physletb.2014.09.022",
    journal = "Phys. Lett. B",
    volume = "738",
    pages = "1--5",
    year = "2014"
}

@article{Frixione:2015zaa,
    author = "Frixione, S. and Hirschi, V. and Pagani, D. and Shao, H. -S. and Zaro, M.",
    title = "{Electroweak and QCD corrections to top-pair hadroproduction in association with heavy bosons}",
    eprint = "1504.03446",
    archivePrefix = "arXiv",
    primaryClass = "hep-ph",
    reportNumber = "CERN-PH-TH-2015-083, CP3-15-10, LPN15-022, SLAC-PUB-16253",
    doi = "10.1007/JHEP06(2015)184",
    journal = "JHEP",
    volume = "06",
    pages = "184",
    year = "2015"
}

@article{Frederix:2018nkq,
    author = "Frederix, R. and Frixione, S. and Hirschi, V. and Pagani, D. and Shao, H. -S. and Zaro, M.",
    title = "{The automation of next-to-leading order electroweak calculations}",
    eprint = "1804.10017",
    archivePrefix = "arXiv",
    primaryClass = "hep-ph",
    reportNumber = "Nikhef/2018-015, TUM-HEP-1138/18, NIKHEF-2018-015, TUM-HEP-1138-18",
    doi = "10.1007/JHEP11(2021)085",
    journal = "JHEP",
    volume = "07",
    pages = "185",
    year = "2018",
    note = "[Erratum: JHEP 11, 085 (2021)]"
}

@article{Catani:2021cbl,
    author = "Catani, Stefano and Fabre, Ignacio and Grazzini, Massimiliano and Kallweit, Stefan",
    title = "{${t {{\bar{t}}}H}$ production at NNLO: the flavour off-diagonal channels}",
    eprint = "2102.03256",
    archivePrefix = "arXiv",
    primaryClass = "hep-ph",
    reportNumber = "ZU-TH 3/21, ICAS 061/21",
    doi = "10.1140/epjc/s10052-021-09247-w",
    journal = "Eur. Phys. J. C",
    volume = "81",
    number = "6",
    pages = "491",
    year = "2021"
}

@article{Catani:2022mfv,
    author = "Catani, Stefano and Devoto, Simone and Grazzini, Massimiliano and Kallweit, Stefan and Mazzitelli, Javier and Savoini, Chiara",
    title = "{Higgs Boson Production in Association with a Top-Antitop Quark Pair in Next-to-Next-to-Leading Order QCD}",
    eprint = "2210.07846",
    archivePrefix = "arXiv",
    primaryClass = "hep-ph",
    reportNumber = "TIF-UNIMI-2022-15, ZU-TH 46/22, MPP-2022-128, PSI-PR-22-30",
    doi = "10.1103/PhysRevLett.130.111902",
    journal = "Phys. Rev. Lett.",
    volume = "130",
    number = "11",
    pages = "111902",
    year = "2023"
}

@article{Devoto:2024nhl,
    author = "Devoto, Simone and Grazzini, Massimiliano and Kallweit, Stefan and Mazzitelli, Javier and Savoini, Chiara",
    title = "{Precise predictions for $ t\overline{t}H $ production at the LHC: inclusive cross section and differential distributions}",
    eprint = "2411.15340",
    archivePrefix = "arXiv",
    primaryClass = "hep-ph",
    reportNumber = "ZU-TH-57/24, PSI-PR-24-22, TUM-HEP-1538/24",
    doi = "10.1007/JHEP03(2025)189",
    journal = "JHEP",
    volume = "03",
    pages = "189",
    year = "2025"
}

@article{FebresCordero:2023pww,
    author = "Febres Cordero, F. and Figueiredo, G. and Kraus, M. and Page, B. and Reina, L.",
    title = "{Two-loop master integrals for leading-color $ pp\to t\overline{t}H $ amplitudes with a light-quark loop}",
    eprint = "2312.08131",
    archivePrefix = "arXiv",
    primaryClass = "hep-ph",
    reportNumber = "CERN-TH-2023-240",
    doi = "10.1007/JHEP07(2024)084",
    journal = "JHEP",
    volume = "07",
    pages = "084",
    year = "2024"
}

@article{Agarwal:2024jyq,
    author = "Agarwal, Bakul and Heinrich, Gudrun and Jones, Stephen P. and Kerner, Matthias and Klein, Sven Yannick and Lang, Jannis and Magerya, Vitaly and Olsson, Anton",
    title = "{Two-loop amplitudes for $ t\overline{t}H $ production: the quark-initiated N$_{f}$-part}",
    eprint = "2402.03301",
    archivePrefix = "arXiv",
    primaryClass = "hep-ph",
    reportNumber = "IPPP/24/03, KA-TP-02-2024, P3H-24-007, TTK-24-03",
    doi = "10.1007/JHEP05(2024)013",
    journal = "JHEP",
    volume = "05",
    pages = "013",
    year = "2024",
    note = "[Erratum: JHEP 06, 142 (2024)]"
}

@article{Wang:2024pmv,
    author = "Wang, Guoxing and Xia, Tianya and Yang, Li Lin and Ye, Xiaoping",
    title = "{Two-loop QCD amplitudes for $ t\overline{t}H $ production from boosted limit}",
    eprint = "2402.00431",
    archivePrefix = "arXiv",
    primaryClass = "hep-ph",
    doi = "10.1007/JHEP07(2024)121",
    journal = "JHEP",
    volume = "07",
    pages = "121",
    year = "2024"
}

@article{Kulesza:2018tqz,
    author = {Kulesza, Anna and Motyka, Leszek and Schwartl{\"a}nder, Daniel and Stebel, Tomasz and Theeuwes, Vincent},
    title = "{Associated production of a top quark pair with a heavy electroweak gauge boson at NLO$+$NNLL accuracy}",
    eprint = "1812.08622",
    archivePrefix = "arXiv",
    primaryClass = "hep-ph",
    reportNumber = "MS-TP-18-29",
    doi = "10.1140/epjc/s10052-019-6746-z",
    journal = "Eur. Phys. J. C",
    volume = "79",
    number = "3",
    pages = "249",
    year = "2019"
}

@article{Kulesza:2017ukk,
    author = "Kulesza, Anna and Motyka, Leszek and Stebel, Tomasz and Theeuwes, Vincent",
    title = "{Associated $t \bar{t} H$ production at the LHC: Theoretical predictions at NLO+NNLL accuracy}",
    eprint = "1704.03363",
    archivePrefix = "arXiv",
    primaryClass = "hep-ph",
    reportNumber = "MS-TP-17-06",
    doi = "10.1103/PhysRevD.97.114007",
    journal = "Phys. Rev. D",
    volume = "97",
    number = "11",
    pages = "114007",
    year = "2018"
}

@article{Kulesza:2015vda,
    author = "Kulesza, Anna and Motyka, Leszek and Stebel, Tomasz and Theeuwes, Vincent",
    title = "{Soft gluon resummation for associated $t \bar{t} H$ production at the LHC}",
    eprint = "1509.02780",
    archivePrefix = "arXiv",
    primaryClass = "hep-ph",
    reportNumber = "MS-TP-15-15",
    doi = "10.1007/JHEP03(2016)065",
    journal = "JHEP",
    volume = "03",
    pages = "065",
    year = "2016"
}

@article{Broggio:2016lfj,
    author = "Broggio, Alessandro and Ferroglia, Andrea and Pecjak, Ben D. and Yang, Li Lin",
    title = "{NNLL resummation for the associated production of a top pair and a Higgs boson at the LHC}",
    eprint = "1611.00049",
    archivePrefix = "arXiv",
    primaryClass = "hep-ph",
    reportNumber = "TUM-HEP-1064-16, IPPP-16-101",
    doi = "10.1007/JHEP02(2017)126",
    journal = "JHEP",
    volume = "02",
    pages = "126",
    year = "2017"
}

@article{Broggio:2015lya,
    author = "Broggio, Alessandro and Ferroglia, Andrea and Pecjak, Ben D. and Signer, Adrian and Yang, Li Lin",
    title = "{Associated production of a top pair and a Higgs boson beyond NLO}",
    eprint = "1510.01914",
    archivePrefix = "arXiv",
    primaryClass = "hep-ph",
    reportNumber = "PSI-PR-15-09, TUM-HEP-1019-15, ZU-TH-33-15, DCPT-15-122, IPPP-15-61",
    doi = "10.1007/JHEP03(2016)124",
    journal = "JHEP",
    volume = "03",
    pages = "124",
    year = "2016"
}

@article{Broggio:2019ewu,
    author = "Broggio, Alessandro and Ferroglia, Andrea and Frederix, Rikkert and Pagani, Davide and Pecjak, Benjamin D. and Tsinikos, Ioannis",
    title = "{Top-quark pair hadroproduction in association with a heavy boson at NLO+NNLL including EW corrections}",
    eprint = "1907.04343",
    archivePrefix = "arXiv",
    primaryClass = "hep-ph",
    reportNumber = "TUM-HEP-1208/19, LU-TP 19-30, IPPP/19/57",
    doi = "10.1007/JHEP08(2019)039",
    journal = "JHEP",
    volume = "08",
    pages = "039",
    year = "2019"
}

@article{Kulesza:2020nfh,
    author = {Kulesza, Anna and Motyka, Leszek and Schwartl{\"a}nder, Daniel and Stebel, Tomasz and Theeuwes, Vincent},
    title = "{Associated top quark pair production with a heavy boson: differential cross sections at NLO+NNLL accuracy}",
    eprint = "2001.03031",
    archivePrefix = "arXiv",
    primaryClass = "hep-ph",
    reportNumber = "MS-TP-20-01",
    doi = "10.1140/epjc/s10052-020-7987-6",
    journal = "Eur. Phys. J. C",
    volume = "80",
    number = "5",
    pages = "428",
    year = "2020"
}

@article{Braaten:2015ppa,
    author = "Braaten, Eric and Zhang, Hong",
    title = "{Inclusive Higgs Production at Large Transverse Momentum}",
    eprint = "1510.01686",
    archivePrefix = "arXiv",
    primaryClass = "hep-ph",
    doi = "10.1103/PhysRevD.93.053014",
    journal = "Phys. Rev. D",
    volume = "93",
    number = "5",
    pages = "053014",
    year = "2016"
}

@article{Mele:1990cw,
    author = "Mele, B. and Nason, P.",
    title = "{The Fragmentation function for heavy quarks in QCD}",
    reportNumber = "CERN-TH-5972-90, UPRF-90-292",
    doi = "10.1016/0550-3213(91)90597-Q",
    journal = "Nucl. Phys. B",
    volume = "361",
    pages = "626--644",
    year = "1991",
    note = "[Erratum: Nucl.Phys.B 921, 841--842 (2017)]"
}

@article{Berman:1971xz,
    author = "Berman, S. M. and Bjorken, J. D. and Kogut, John B.",
    title = "{Inclusive Processes at High Transverse Momentum}",
    reportNumber = "SLAC-PUB-0944",
    doi = "10.1103/PhysRevD.4.3388",
    journal = "Phys. Rev. D",
    volume = "4",
    pages = "3388",
    year = "1971"
}

@article{Gao:2025hlm,
    author = "Gao, Jun and Shen, XiaoMin and Xing, Hongxi and Zhao, Yuxiang and Zhou, Bin",
    title = "{Fragmentation Functions of Charged Hadrons at Next-to-Next-to-Leading Order and Constraints on the Proton Parton Distribution Functions}",
    eprint = "2502.17837",
    archivePrefix = "arXiv",
    primaryClass = "hep-ph",
    doi = "10.1103/mcwy-b221",
    journal = "Phys. Rev. Lett.",
    volume = "135",
    number = "4",
    pages = "041902",
    year = "2025"
}

@article{AbdulKhalek:2022laj,
    author = "Abdul Khalek, Rabah and Bertone, Valerio and Khoudli, Alice and Nocera, Emanuele R.",
    collaboration = "MAP (Multi-dimensional Analyses of Partonic distributions)",
    title = "{Pion and kaon fragmentation functions at next-to-next-to-leading order}",
    eprint = "2204.10331",
    archivePrefix = "arXiv",
    primaryClass = "hep-ph",
    reportNumber = "JLAB-THY-22-3605",
    doi = "10.1016/j.physletb.2022.137456",
    journal = "Phys. Lett. B",
    volume = "834",
    pages = "137456",
    year = "2022"
}

@article{Moffat:2021dji,
    author = "Moffat, Eric and Melnitchouk, Wally and Rogers, T. C. and Sato, Nobuo",
    collaboration = "Jefferson Lab Angular Momentum (JAM)",
    title = "{Simultaneous Monte~Carlo analysis of parton densities and fragmentation functions}",
    eprint = "2101.04664",
    archivePrefix = "arXiv",
    primaryClass = "hep-ph",
    reportNumber = "JLAB-THY-21-3304",
    doi = "10.1103/PhysRevD.104.016015",
    journal = "Phys. Rev. D",
    volume = "104",
    number = "1",
    pages = "016015",
    year = "2021"
}

@article{Aivazis:1993kh,
    author = "Aivazis, M. A. G. and Olness, Frederick I. and Tung, Wu-Ki",
    title = "{Leptoproduction of heavy quarks. 1. General formalism and kinematics of charged current and neutral current production processes}",
    eprint = "hep-ph/9312318",
    archivePrefix = "arXiv",
    reportNumber = "MSUHEP-93-15, SMU-HEP-93-16",
    doi = "10.1103/PhysRevD.50.3085",
    journal = "Phys. Rev. D",
    volume = "50",
    pages = "3085--3101",
    year = "1994"
}

@article{Aivazis:1993pi,
    author = "Aivazis, M. A. G. and Collins, John C. and Olness, Fredrick I. and Tung, Wu-Ki",
    title = "{Leptoproduction of heavy quarks. 2. A Unified QCD formulation of charged and neutral current processes from fixed target to collider energies}",
    eprint = "hep-ph/9312319",
    archivePrefix = "arXiv",
    reportNumber = "SMU-HEP-93-17, MSUHEP-93-17, PSU-TH-138",
    doi = "10.1103/PhysRevD.50.3102",
    journal = "Phys. Rev. D",
    volume = "50",
    pages = "3102--3118",
    year = "1994"
}

@article{Kramer:2000hn,
    author = {Kr{\"a}mer, Michael and Olness, Fredrick I. and Soper, Davison E.},
    title = "{Treatment of heavy quarks in deeply inelastic scattering}",
    eprint = "hep-ph/0003035",
    archivePrefix = "arXiv",
    reportNumber = "EDINBURGH-2000-02",
    doi = "10.1103/PhysRevD.62.096007",
    journal = "Phys. Rev. D",
    volume = "62",
    pages = "096007",
    year = "2000"
}

@article{Cacciari:1998it,
    author = "Cacciari, Matteo and Greco, Mario and Nason, Paolo",
    title = "{The $p_T$ spectrum in heavy-flavour hadroproduction.}",
    eprint = "hep-ph/9803400",
    archivePrefix = "arXiv",
    reportNumber = "CERN-TH-98-77, LPTHE-ORSAY-98-11, IFUM-613-FT, LNF-98-008-P",
    doi = "10.1088/1126-6708/1998/05/007",
    journal = "JHEP",
    volume = "05",
    pages = "007",
    year = "1998"
}

@article{Buza:1996wv,
    author = "Buza, M. and Matiounine, Y. and Smith, J. and van Neerven, W. L.",
    title = "{Charm electroproduction viewed in the variable flavor number scheme versus fixed order perturbation theory}",
    eprint = "hep-ph/9612398",
    archivePrefix = "arXiv",
    reportNumber = "NIKHEF-96-027, ITP-SB-96-66, DESY-96-258, INLO-PUB-22-96",
    doi = "10.1007/BF01245820",
    journal = "Eur. Phys. J. C",
    volume = "1",
    pages = "301--320",
    year = "1998"
}

@article{NNPDF:2021njg,
    author = "Ball, Richard D. and others",
    collaboration = "NNPDF",
    title = "{The path to proton structure at 1{\%} accuracy}",
    eprint = "2109.02653",
    archivePrefix = "arXiv",
    primaryClass = "hep-ph",
    reportNumber = "Edinburgh 2021/12, Nikhef-2021-013, TIF-UNIMI-2021-11",
    doi = "10.1140/epjc/s10052-022-10328-7",
    journal = "Eur. Phys. J. C",
    volume = "82",
    number = "5",
    pages = "428",
    year = "2022"
}

@article{Generet:2025gdy,
    author = "Generet, Terry",
    title = "{IRC-safe jet flavour without modifying anything}",
    eprint = "2511.23423",
    archivePrefix = "arXiv",
    primaryClass = "hep-ph",
    reportNumber = "Cavendish-HEP-25/06",
    month = "11",
    year = "2025"
}

@article{Buccioni:2019sur,
    author = {Buccioni, Federico and Lang, Jean-Nicolas and Lindert, Jonas M. and Maierh{\"o}fer, Philipp and Pozzorini, Stefano and Zhang, Hantian and Zoller, Max F.},
    title = "{OpenLoops 2}",
    eprint = "1907.13071",
    archivePrefix = "arXiv",
    primaryClass = "hep-ph",
    reportNumber = "IPPP/19/62, FR-PHENO-2019-12, PSI-PR-19-15, ZU-TH 37/19",
    doi = "10.1140/epjc/s10052-019-7306-2",
    journal = "Eur. Phys. J. C",
    volume = "79",
    number = "10",
    pages = "866",
    year = "2019"
}

@article{Czakon:2014oma,
    author = "Czakon, M. and Heymes, D.",
    title = "{Four-dimensional formulation of the sector-improved residue subtraction scheme}",
    eprint = "1408.2500",
    archivePrefix = "arXiv",
    primaryClass = "hep-ph",
    reportNumber = "TTK-14-16",
    doi = "10.1016/j.nuclphysb.2014.11.006",
    journal = "Nucl. Phys. B",
    volume = "890",
    pages = "152--227",
    year = "2014"
}

@article{Czakon:2010td,
    author = "Czakon, M.",
    title = "{A novel subtraction scheme for double-real radiation at NNLO}",
    eprint = "1005.0274",
    archivePrefix = "arXiv",
    primaryClass = "hep-ph",
    doi = "10.1016/j.physletb.2010.08.036",
    journal = "Phys. Lett. B",
    volume = "693",
    pages = "259--268",
    year = "2010"
}

@article{Czakon:2011ve,
    author = "Czakon, M.",
    title = "{Double-real radiation in hadronic top quark pair production as a proof of a certain concept}",
    eprint = "1101.0642",
    archivePrefix = "arXiv",
    primaryClass = "hep-ph",
    reportNumber = "TTK-10-58, SFB-CPP-10-134",
    doi = "10.1016/j.nuclphysb.2011.03.020",
    journal = "Nucl. Phys. B",
    volume = "849",
    pages = "250--295",
    year = "2011"
}

@article{Czakon:2019tmo,
    author = "Czakon, Micha{\l} and van Hameren, Andreas and Mitov, Alexander and Poncelet, Rene",
    title = "{Single-jet inclusive rates with exact color at $ \mathcal{O} $ ($ {\alpha}_s^4 $)}",
    eprint = "1907.12911",
    archivePrefix = "arXiv",
    primaryClass = "hep-ph",
    doi = "10.1007/JHEP10(2019)262",
    journal = "JHEP",
    volume = "10",
    pages = "262",
    year = "2019"
}

@article{Vermaseren:2000nd,
    author = "Vermaseren, J. A. M.",
    title = "{New features of FORM}",
    eprint = "math-ph/0010025",
    archivePrefix = "arXiv",
    month = "10",
    year = "2000"
}

@article{Kuipers:2012rf,
    author = "Kuipers, J. and Ueda, T. and Vermaseren, J. A. M. and Vollinga, J.",
    title = "{FORM version 4.0}",
    eprint = "1203.6543",
    archivePrefix = "arXiv",
    primaryClass = "cs.SC",
    reportNumber = "NIKHEF-2012-004, TTP12-008, SFB-CPP-12-15",
    doi = "10.1016/j.cpc.2012.12.028",
    journal = "Comput. Phys. Commun.",
    volume = "184",
    pages = "1453--1467",
    year = "2013"
}

@article{Buckley:2014ana,
    author = {Buckley, Andy and Ferrando, James and Lloyd, Stephen and Nordstr{\"o}m, Karl and Page, Ben and R{\"u}fenacht, Martin and Sch{\"o}nherr, Marek and Watt, Graeme},
    title = "{LHAPDF6: parton density access in the LHC precision era}",
    eprint = "1412.7420",
    archivePrefix = "arXiv",
    primaryClass = "hep-ph",
    reportNumber = "GLAS-PPE-2014-05, MCNET-14-29, IPPP-14-111, DCPT-14-222",
    doi = "10.1140/epjc/s10052-015-3318-8",
    journal = "Eur. Phys. J. C",
    volume = "75",
    pages = "132",
    year = "2015"
}

@article{Biello:2026nhj,
    author = "Biello, Christian and Savoini, Chiara and Signorile-Signorile, Chiara and Wiesemann, Marius",
    title = "{Next-to-next-to-leading order event generation for $t\bar{t}H$ production with approximate two-loop amplitude}",
    eprint = "2603.06143",
    archivePrefix = "arXiv",
    primaryClass = "hep-ph",
    reportNumber = "CERN-TH-2026-031, MPP-2026-22, TUM-HEP-1595/26",
    month = "3",
    year = "2026"
}

@article{Kinoshita:1962ur,
    author = "Kinoshita, T.",
    title = "{Mass singularities of Feynman amplitudes}",
    doi = "10.1063/1.1724268",
    journal = "J. Math. Phys.",
    volume = "3",
    pages = "650--677",
    year = "1962"
}

@article{Lee:1964is,
    author = "Lee, T. D. and Nauenberg, M.",
    editor = "Feinberg, G.",
    title = "{Degenerate Systems and Mass Singularities}",
    doi = "10.1103/PhysRev.133.B1549",
    journal = "Phys. Rev.",
    volume = "133",
    pages = "B1549--B1562",
    year = "1964"
}

@article{Buckley:2021gfw,
    author = "Buckley, A. and others",
    title = "{A comparative study of Higgs boson production from vector-boson fusion}",
    eprint = "2105.11399",
    archivePrefix = "arXiv",
    primaryClass = "hep-ph",
    reportNumber = "FERMILAB-PUB-21-218-T, IPPP/20/101, MCNET-21-08, KA-TP-08-2021, KA-TP-08-2021,
  OUTP-21-14P, ZU-TH 22/21, CERN-TH-2021-081, OUTP-21-14P",
    doi = "10.1007/JHEP11(2021)108",
    journal = "JHEP",
    volume = "11",
    pages = "108",
    year = "2021"
}

@article{Manohar:2006nz,
    author = "Manohar, Aneesh V. and Stewart, Iain W.",
    title = "{The Zero-Bin and Mode Factorization in Quantum Field Theory}",
    eprint = "hep-ph/0605001",
    archivePrefix = "arXiv",
    reportNumber = "MIT-CTP-3726, UCSD-PTH-06-04",
    doi = "10.1103/PhysRevD.76.074002",
    journal = "Phys. Rev. D",
    volume = "76",
    pages = "074002",
    year = "2007"
}

@article{Catani:2011st,
    author = "Catani, Stefano and de Florian, Daniel and Rodrigo, German",
    title = "{Space-like (versus time-like) collinear limits in QCD: Is factorization violated?}",
    eprint = "1112.4405",
    archivePrefix = "arXiv",
    primaryClass = "hep-ph",
    reportNumber = "LPN11-94, IFIC-11-72, ZU-TH-27-11",
    doi = "10.1007/JHEP07(2012)026",
    journal = "JHEP",
    volume = "07",
    pages = "026",
    year = "2012"
}

@article{Forshaw:2012bi,
    author = "Forshaw, Jeffrey R. and Seymour, Michael H. and Siodmok, Andrzej",
    title = "{On the Breaking of Collinear Factorization in QCD}",
    eprint = "1206.6363",
    archivePrefix = "arXiv",
    primaryClass = "hep-ph",
    reportNumber = "MAN-HEP-2012-05",
    doi = "10.1007/JHEP11(2012)066",
    journal = "JHEP",
    volume = "11",
    pages = "066",
    year = "2012"
}

@article{Schwartz:2017nmr,
    author = "Schwartz, Matthew D. and Yan, Kai and Zhu, Hua Xing",
    title = "{Collinear factorization violation and effective field theory}",
    eprint = "1703.08572",
    archivePrefix = "arXiv",
    primaryClass = "hep-ph",
    doi = "10.1103/PhysRevD.96.056005",
    journal = "Phys. Rev. D",
    volume = "96",
    number = "5",
    pages = "056005",
    year = "2017"
}

@article{Henn:2024qjq,
    author = "Henn, Johannes and Ma, Rourou and Xu, Yongqun and Yan, Kai and Zhang, Yang and Zhu, Hua Xing",
    title = "{Two-loop spacelike splitting amplitude for N=4 Super-Yang-Mills theory}",
    eprint = "2406.14604",
    archivePrefix = "arXiv",
    primaryClass = "hep-th",
    reportNumber = "USTC-ICTS/PCFT-24-18",
    doi = "10.1103/5qrc-4vq1",
    journal = "Phys. Rev. D",
    volume = "112",
    number = "7",
    pages = "076003",
    year = "2025"
}

@article{Guan:2024hlf,
    author = "Guan, Xin and Herzog, Franz and Ma, Yao and Mistlberger, Bernhard and Suresh, Adi",
    title = "{Splitting amplitudes at N$^{3}$LO in QCD}",
    eprint = "2408.03019",
    archivePrefix = "arXiv",
    primaryClass = "hep-ph",
    reportNumber = "SLAC-PUB-17781",
    doi = "10.1007/JHEP01(2025)090",
    journal = "JHEP",
    volume = "01",
    pages = "090",
    year = "2025"
}

@article{Weinberg:1980wa,
    author = "Weinberg, Steven",
    title = "{Effective Gauge Theories}",
    reportNumber = "HUTP-80/A001",
    doi = "10.1016/0370-2693(80)90660-7",
    journal = "Phys. Lett. B",
    volume = "91",
    pages = "51--55",
    year = "1980"
}

@article{Ovrut:1980dg,
    author = "Ovrut, Burt A. and Schnitzer, Howard J.",
    title = "{The Decoupling Theorem and Minimal Subtraction}",
    reportNumber = "Print-80-0659 (BRANDEIS)",
    doi = "10.1016/0370-2693(81)90146-5",
    journal = "Phys. Lett. B",
    volume = "100",
    pages = "403--406",
    year = "1981"
}

@article{Buza:1995ie,
    author = "Buza, M. and Matiounine, Y. and Smith, J. and Migneron, R. and van Neerven, W. L.",
    title = "{Heavy quark coefficient functions at asymptotic values Q**2 {\ensuremath{>}}{\ensuremath{>}} m**2}",
    eprint = "hep-ph/9601302",
    archivePrefix = "arXiv",
    reportNumber = "NIKHEF-95-070, ITP-SB-95-59, INLO-PUB-22-95",
    doi = "10.1016/0550-3213(96)00228-3",
    journal = "Nucl. Phys. B",
    volume = "472",
    pages = "611--658",
    year = "1996"
}

@article{Bierenbaum:2009zt,
    author = "Bierenbaum, Isabella and Blumlein, Johannes and Klein, Sebastian",
    title = "{The Gluonic Operator Matrix Elements at O(alpha(s)**2) for DIS Heavy Flavor Production}",
    eprint = "0901.0669",
    archivePrefix = "arXiv",
    primaryClass = "hep-ph",
    reportNumber = "DESY-08-187, SFB-CPP-08-107, IFIC-08-68",
    doi = "10.1016/j.physletb.2009.01.057",
    journal = "Phys. Lett. B",
    volume = "672",
    pages = "401--406",
    year = "2009"
}

@article{Braaten:1980yq,
    author = "Braaten, E. and Leveille, J. P.",
    title = "{Higgs Boson Decay and the Running Mass}",
    reportNumber = "COO-881-127",
    doi = "10.1103/PhysRevD.22.715",
    journal = "Phys. Rev. D",
    volume = "22",
    pages = "715",
    year = "1980"
}

@article{Tarrach:1980up,
    author = "Tarrach, R.",
    title = "{The Pole Mass in Perturbative QCD}",
    reportNumber = "CPT-80/P-1215",
    doi = "10.1016/0550-3213(81)90140-1",
    journal = "Nucl. Phys. B",
    volume = "183",
    pages = "384--396",
    year = "1981"
}

@article{Dawson:1997im,
    author = "Dawson, S. and Reina, L.",
    title = "{QCD corrections to associated Higgs boson production}",
    eprint = "hep-ph/9712400",
    archivePrefix = "arXiv",
    reportNumber = "MADPH-97-1027",
    doi = "10.1103/PhysRevD.57.5851",
    journal = "Phys. Rev. D",
    volume = "57",
    pages = "5851--5859",
    year = "1998"
}

\end{document}